\documentclass{aastex631}

\newcommand{\RomanNumeralCaps}[1]
{\MakeUppercase{\romannumeral #1}}
\newcommand{\microns}[1]{{#1} $\mu$m}

\newcommand{\rom}[1]{\uppercase\expandafter{\romannumeral#1}}

\accepted{to the Astrophysical Journal April 07, 2026}

\begin{document}

\title{EPISODE IV: Ice Inventory in the Envelope of EC 53}

\author[0000-0001-8604-2801]{Jaeyeong Kim}
\affil{Korea Astronomy and Space Science Institute 776 Daedeok-daero, Yuseong-gu, Daejeon 34055, Republic of Korea}
\affiliation{Center for Astrophysics $|$ Harvard \& Smithsonian, 60 Garden Street, Cambridge, MA 02138-1516; USA}
\email{jaeyeong@kasi.re.kr} 

\author[0000-0003-3119-2087]{Jeong-Eun Lee}
\affil{Department of Physics and Astronomy, Seoul National University, 1 Gwanak-ro, Gwanak-gu, Seoul 08826, Republic of Korea}
\affil{SNU Astronomy Research Center, Seoul National University, 1 Gwanak-ro, Gwanak-gu, Seoul 08826, Republic of Korea}
\correspondingauthor{Jeong-Eun Lee}
\email{lee.jeongeun@snu.ac.kr} 

\author[0000-0002-2523-3762]{Chul-Hwan Kim}
\affil{Department of Physics and Astronomy, Seoul National University, 1 Gwanak-ro, Gwanak-gu, Seoul 08826, Republic of Korea}

\author[0000-0002-0226-9295]{Seokho Lee}
\affil{Korea Astronomy and Space Science Institute 776 Daedeok-daero, Yuseong-gu, Daejeon 34055, Republic of Korea}

\author[0000-0002-2814-1978]{Giseon Baek}
\affil{Department of Physics and Astronomy, Seoul National University, 1 Gwanak-ro, Gwanak-gu, Seoul 08826, Republic of Korea}
\affil{Research Institute of Basic Sciences, Seoul National University, Seoul 08826, Republic of Korea}

\author[0000-0002-2523-3762]{Seonjae Lee}
\affil{Department of Physics and Astronomy, Seoul National University, 1 Gwanak-ro, Gwanak-gu, Seoul 08826, Republic of Korea}

\author[0000-0001-8227-2816]{Yao-Lun Yang}
\affil{RIKEN Pioneering Research Institute, Wako-shi, Saitama, 351-0198, Japan}

\author[0000-0003-3283-6884]{Yuri Aikawa}
\affil{Department of Astronomy, University of Tokyo, 7-3-1 Hongo, Bunkyo-ku, Tokyo 113-0033, Japan}

\author[0000-0002-7154-6065]{Gregory J. Herczeg}
\affil{Kavli Institute for Astronomy and Astrophysics, Peking University, Yiheyuan Lu 5, Haidian Qu, 100871, Beijing, PR China}

\author[0000-0002-6773-459X]{Doug Johnstone}
\affil{NRC Herzberg Astronomy and Astrophysics, 5071 West Saanich Rd, Victoria, BC, V9E 2E7, Canada }
\affil{Department of Physics and Astronomy, University of Victoria, Victoria, BC, V8P 5C2, Canada}

\author[0000-0003-1665-5709]{Joel D. Green}
\affil{Space Telescope Science Institute, 3700 San Martin Dr., Baltimore, MD 02138, USA}

\author[0000-0002-2770-808X]{Woong-Seob Jeong}
\affil{Korea Astronomy and Space Science Institute 776 Daedeok-daero, Yuseong-gu, Daejeon 34055, Republic of Korea}




\begin{abstract}
We present the 1.6--\microns{28} spectra of the young protostar EC 53, obtained with JWST NIRSpec IFU and MIRI MRS during the quiescent and burst phases of its periodic brightness variations.
To isolate ice absorption features, we modeled and removed the mid-infrared silicate dust absorption using a dedicated continuum-fitting procedure. 
In the optical depth spectrum, we first fit the broad H$_2$O ice features and then decomposed the major ice components, including NH$_3$, CO$_2$, CH$_3$OH, CO, and CH$_4$, by matching laboratory profiles for both pure and H$_2$O-mixed ices. The \microns{4.62} and \microns{6.85} bands are attributed to OCN$^-$ and NH$_4^+$ ions, respectively. 
Minor or tentative contributions from complex species (HCOOH, H$_2$CO, CH$_3$COOH, CH$_3$CHO, CH$_3$CH$_2$OH, and NH$_2$CHO ) are also considered to our global ice analysis.
The silicate-corrected spectra reveal no measurable change in any ice absorption band between the two phases, indicating that moderate and short-period accretion bursts in EC~53 do not significantly alter the physical or chemical state of the ices within its envelope.
The derived abundances of these major species relative to H$_2$O significantly exceed the values typically observed toward other embedded protostars.
Finally, we place the ice inventory of EC 53 in the context of other protostellar systems observed with JWST, highlighting that its chemically rich, thermally quiescent ice reservoir provides a benchmark for studying ice evolution under episodic accretion.

\end{abstract}

\section{Introduction} \label{sec:intro}
Ice chemistry plays a crucial role in the early stages of star formation, setting the initial chemical conditions that govern the evolution of protostellar envelopes and disks. 
In the cold and dense regions of molecular clouds, icy mantles form on dust grains, providing unique sites where simple species react to form increasingly complex molecules \citep{Allamandola1999,Boogert2015}. 
These mantles both record the chemical history of their natal environments and supply material that eventually participates in planet formation. 
In particular, many complex organic molecules (COMs) are believed to form efficiently on grain surfaces within prestellar cores under low-temperature, high-density conditions \citep{Herbst2009,Oberg2016}. 
As protostars begin to accrete, thermal heating can desorb these mantles and trigger additional gas-phase reactions, further enhancing or altering COMs formation pathways \citep{Taquet2016}.

Such chemical evolution, however, is not governed solely by quiescent physical conditions.
Protostellar growth is now understood to proceed through highly variable mass accretion rates rather than steady inflow \citep{Hartmann1998, Dunham2008, Francis2022, Fischer2023}.
Variability in protostellar luminosity driven by episodic accretion can intermittently heat the circumstellar environment, dramatically altering the structure and composition of the ice reservoir.
The Spitzer c2d survey \citep{Evans2009} demonstrated that most protostars exhibit luminosities lower than predicted by steady state models \citep{Enoch2009,Hsieh2013,dunham2014}, implying long periods of low accretion punctuated by short, intense bursts. 
Although several mechanisms, including thermal instabilities \citep{Bell1994}, gravitational instabilities \citep{Vorobyov2005,Vorobyov2010,Machida2011}, and magneto-rotational instabilities \citep{Armitage2001,Kadam2020,Cleaver2023}, have been proposed to explain these triggers, their relative contributions remain debated. 
Simulations suggest that inward migration of gravitationally unstable disk fragments may drive bursts during the Class 0/I phases \citep{Vorobyov2015}, consistent with observations showing a higher burst frequency in Class 0 compared to Class I sources \citep{Hsieh2018,Hsieh2019}.
 

Accretion bursts are expected to strongly affect ice chemistry by heating the envelope, activating high-energy chemical processes, annealing, and sublimation of ice mantles \citep{JLee2007,Cieza2016,Molyarova2018,Wiebe2019}. Even after cooling, chemical imprints of past bursts persist in the ice mantles, offering a time-resolved record of episodic heating. Infrared spectroscopic observations with Spitzer and AKARI revealed these chemical imprints, such as double-peaked CO$_2$ ice features at \microns{15.2} \citep{Kim2012} and comparative CO$_2$/CO ice abundances toward low-luminosity protostars \citep{Kim2022}.
However, we still lack direct, time-resolved measurements of how ices respond during luminosity variations.

The advent of the James Webb Space Telescope \citep[JWST;][]{Gardner2006}, featuring unprecedented sensitivity and broad spectral coverage, has significantly expanded our ability to probe ice compositions and formation pathways in dense regions across molecular clouds and protostellar environments \citep{Yang2022,McClure2023,Smith2025}.
Large programs such as JOYS, IPA, and IceAge have delivered comprehensive ice inventories, discovered rare species, and detected potential COMs \citep{chen2024,rocha2024,Rocha2025,Brunken2024a,Brunken2024b,Nazari2024,Slavicinska2024,vanDishoeck2025}. 
However, these broad surveys primarily provide single-epoch snapshots and therefore cannot directly link measured ice compositions to specific accretion histories.
Because the timescales for ice sublimation and freeze-out differ substantially in response to luminosity variations, understanding how episodic accretion affects ice chemistry requires targeted, phase-resolved observations of sources with predictable luminosity changes.

EC 53 (V371 Ser), an embedded Class I protostar in Serpens Main \citep[d=436~pc,][]{Ortiz-Leon2017}, provides a rare opportunity for such a study. It exhibits quasi-periodic near-infrared-to-submillimeter brightness variations with a $\sim$1.5-year cycle \citep{Hodapp1999,Hodapp2012,YLee2020,Francis2022}. These regular bursts, which are potentially driven by disk–companion interactions at a few au \citep{Bonnell1992,Nayakshin2012,Baek2020}, provide a natural laboratory for studying how rapid luminosity changes affect ice sublimation, re-freeze, and chemical evolution.
An ALMA study revealing extended CH$_3$OH emission beyond the current sublimation radius suggests that EC~53 has experienced stronger past bursts \citep{SLee2020}. However, its relatively low gas-phase CH$_3$OH abundance, compared to the outbursting disk of V883~Ori \citep{JLee2019}, may indicate rapid re-freeze in its dense inner envelope. In contrast, B335, which recently underwent an accretion burst, shows that the freeze-out timescale can be significantly longer than expected theoretically, likely due to prolonged cooling in its high-density inner region \citep{JLee2025}. The high sensitivity of JWST to weak ice features makes EC~53 uniquely suited to tracing the short-timescale chemical response of ices throughout its accretion cycle.

In this work, we present JWST NIRSpec IFU and MIRI MRS spectra of EC~53 spanning 1.6$-$\microns{28}, obtained during its quiescent and burst phases, via the JWST GO2 program, {\it EPISODE: EC 53, the only known Periodically variable Infant Star to chase the Outburst in the next Dynamical Event}. 
Focusing on near- and mid-infrared absorption features, we extract and compare the ice compositions present during each phase.
We apply a global ice-decomposition methodology to quantify the major ice species and assess how episodic accretion influences the physical and chemical structure of ices within the EC~53 envelope.
Section~\ref{sec:observations} describes the observations and data reduction; Section~\ref{sec:spectral_analysis} presents spectral characteristics and a continuum-fitting method; Section~\ref{sec:results} details our ice-fitting methodology; Section~\ref{sec:discussions} discusses the chemical implications in the context of other protostars; and Section~\ref{sec:summary} summarizes our conclusions.
Companion papers will separately explore other phase-dependent phenomena using the same dataset: the newly emerging crystalline silicate emission during the burst phase \citep[][Paper I]{JLee2026}, the veiling effect on the CO fundamental band near \microns{4.7} and the H$_2$O rovibrational absorption lines around \microns{6.0} during the burst phase (Seokho Lee et al., submitted; Paper II), and the fast atomic (e.g., [Fe II], [S I], [Ne II]) jets nested by the slower H$_2$ and CO emission, which is evidence of MHD disk-wind driven outflows \citep[][Paper III]{SJLee2026}.

\section{Observations \label{sec:observations}}
EC 53 was observed with JWST by the General Observer (GO) program 3477 (PI: Jeong-Eun Lee). Time-constrained observations were conducted during its quiescent and burst phases on October 5, 2023 and May 10, 2024, respectively. 
The observations employed both the Near Infrared Spectrograph Integral Field Unit \citep[NIRSpec IFU;][]{Boker2022,Jakobsen2022} and MIRI Medium Resolution Spectroscopy \citep[MRS;][]{Rieke2015,Wells2015,Wright2015,Argyriou2023,Wright2023}, covering a broad wavelength range from 1.6 to \microns{28}, in which various molecular ice species exhibit distinct absorption features.

For NIRSpec IFU observations, we used the disperser-filter combinations G235H-F170LP (1.66$-$\microns{3.05}) and G395H-F290LP (2.87$-$\microns{5.14}) to achieve the highest available spectral resolving power (R$\sim$2700).
For MIRI MRS mode, the observations covered a wavelength range of 4.9$-$\microns{27.9} with spectral resolutions ranging from approximately R$\sim$3700 at the shortest wavelengths to R$\sim$1300 at the longest wavelengths. 
The MRS consists of four spectral channels, each further divided into three sub-bands: Short (A), Medium (B), and Long (C).

These observations enable a detailed comparative study of changes in the absorption properties of ice and gas species between the quiescent and burst phases of EC 53. Additional instrumental details and the complete data reduction processes are presented in companion papers: Paper I for MIRI observations and Paper II for NIRSpec observations.
\par
The spectra used in this study were extracted from circular apertures for both the quiescent and burst observations. For the NIRSpec IFU data, the aperture diameter was fixed at 1.3\arcsec.
For the spectral channels in the MIRI MRS data, the aperture diameters were set to four times the FWHM of the MIRI MRS point spread function (PSF) \citep{Law2023}. However, in the short-wavelength channels in MIRI CH1, where the aperture diameters were smaller than 1.3\arcsec, we fixed the aperture diameter to 1.3\arcsec\, to maintain consistency with the NIRSpec IFU spectra. In addition, because the spectral coverage of the two NIRSpec disperser-filter combinations overlaps in a certain wavelength range, we manually trimmed the G235H-F170LP spectrum at the short-wavelength end of the G395-F290LP spectrum.  For the MIRI spectrum, we applied the {\tt fit\_residual\_fringes\_1d} in the JWST pipeline to each sub-band spectrum to correct the fringe patterns. Since the sub-band spectra overlap in a certain wavelength range, we constructed a single MIRI spectrum by averaging the flux in the overlapping wavelength regions.

\section{Spectral analysis\label{sec:spectral_analysis}}
\begin{figure}[htp]
 \centering
 \vspace{-2mm}
 \includegraphics[width=1.0\textwidth]{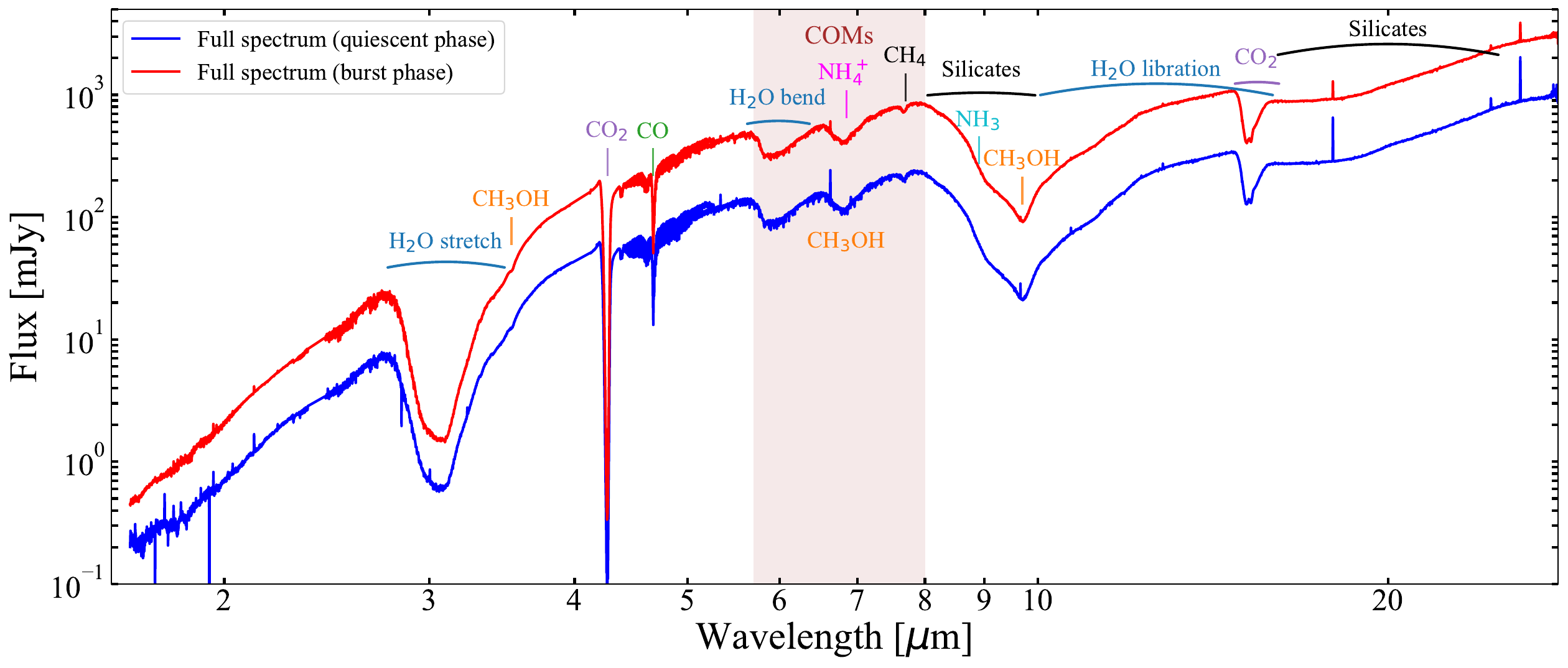}
 \vspace{-5mm}
 \caption{Total spectrum of EC 53 reduced from the NIRSpec IFU and MIRI MRS observations at the quiescent (blue) and burst (red) phases, respectively. Silicate and ice absorption bands are labeled. The complex absorption band of COM ices is highlighted in brown.}
\label{fig:Full_spec_both_epochs}
\end{figure}

Figure~\ref{fig:Full_spec_both_epochs} shows the resulting spectrum of EC 53 in the quiescent and burst phases, obtained from the JWST NIRSpec and MIRI MRS observations described above. 
The flux observed during the burst phase tends to increase by at least a factor of three relative to the quiescent phase, consistent with the previously reported luminosity ratio \citep{Baek2020}.
Both NIRSpec and MIRI spectra exhibit rich spectral features, including deep absorption bands from major ice species such as water (H$_2$O), carbon dioxide ($^{12}$CO$_2$, hereafter CO$_2$), and carbon monoxide (CO).
These are consistently detected in both phases, along with minor features attributed to more complex species.

\subsection{NIRSpec IFU}
While the direct phase-dependent comparison reveals notable changes in the depths of gaseous absorption features (Paper II), variations in the ice absorption features are not readily apparent and require a more precise analysis.
In particular, reliable measurement of ice optical depths requires a careful determination of the underlying continuum level across the full infrared spectral range from NIRSpec to MIRI.
Because the continuum placement establishes the zero-level baseline for optical depth calculations, any errors in the baseline can directly affect the derived ice column densities.
We initially adopted the standard approach commonly used in near-infrared spectral analyses, in which the continuum is approximated by a low-order polynomial fitted to selected spectral bands that avoid strong absorption and emission features \citep{Brooke1999,Gibb2004,Aikawa2012,Kim2022}.
A fourth-order polynomial was fitted to the NIRSpec IFU spectra using manually selected wavelength intervals. 
Although this polynomial fit reproduces the overall continuum curve reasonably well between \microns{2.0} and \microns{5.3}, it fails at shorter wavelengths.
In particular, the steep flux level decrease shortward of \microns{2.0} causes the polynomial fit to diverge from the observed spectral feature, leading to an unreliable continuum estimate in this region.

\begin{figure}[htp]
 \centering
 \vspace{-2mm}
 \includegraphics[width=1.0\textwidth]{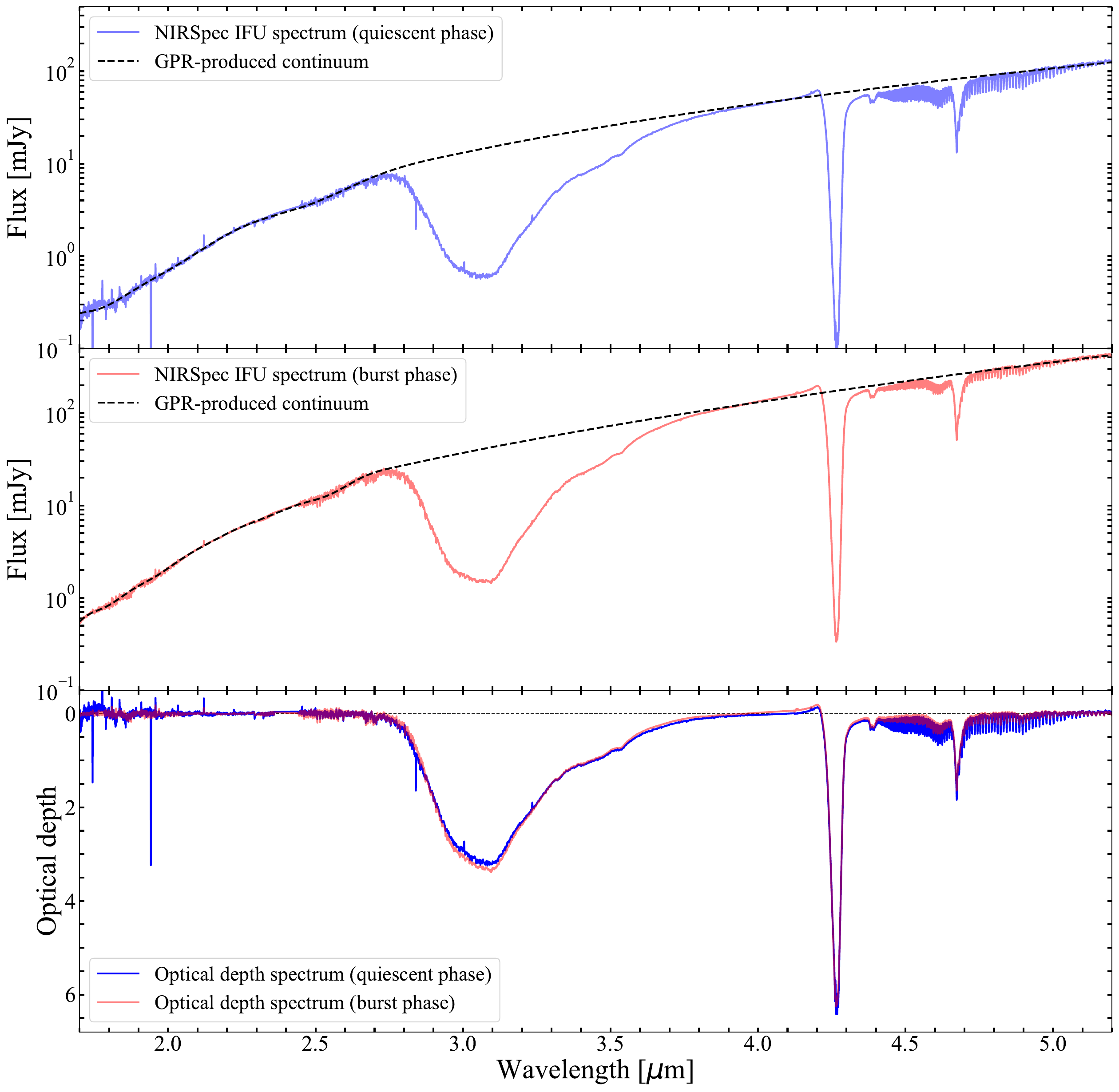}
 \caption{Top: Optimized continuum curve (black dashed line) produced by the Gaussian Process Regression (GPR) method to the NIRSpec spectrum of EC~53 in the quiescent phase. 
 Middle: The same figure of the top panel, but for the burst phase.
 Bottom: Converted optical depth spectrum for the ice absorption features of both phases.}
\label{fig:Cont_fit_NIRSpec_both_epochs}
\end{figure}

To address this limitation and obtain a more robust continuum determination across the full 1.66–\microns{5.27} wavelength range, we implemented a Gaussian Process Regression (GPR) approach with a squared-exponential kernel. 
Gaussian processes provide a non-parametric Bayesian framework for modeling smooth functions.
Rather than assuming a fixed functional form, such as a polynomial, GPR treats the continuum as a stochastic function whose values at different wavelengths are corrected according to a covariance kernel.
The squared-exponential kernel adopted here assumes that the continuum varies smoothly with wavelengths, such that spectral points closer in wavelengths are more strongly correlated than those farther apart.
Given a set of training data points that represent the continuum, the GPR model predicts the most probable underlying function, together with a wavelength-dependent uncertainty estimate, while also providing a probabilistic estimate of the interpolation.

We employed the GPR implementation available in {\tt scikit-learn} \citep[sklearn.gaussian\_process.GaussianProcessRegressor;][]{scikit-learn}.
The training data for the GPR model were selected from spectral intervals devoid of both strong emission and absorption features: 1.66–1.72, 1.78–1.86, 1.88–1.95, 2.03–2.11, 2.13–2.35, 2.40–2.45, and 2.48–\microns{2.64}.
These regions were manually inspected and verified to avoid contamination from known molecular absorption bands or atomic lines.
The GPR then interpolates the continuum across the entire spectral range by exploiting the smooth correlations imposed by the kernel.
Since the broad H$_2$O ice bands near 3.0 and \microns{4.5} dominate the spectrum, we further stabilized the GPR solution by seeding the model with continuum points derived from the polynomial fit in the 2.7–4.0 and 4.1–\microns{5.27} regions.
This hybrid scheme preserves the global spectral trend provided by the polynomial fit in ice-dominated regions while allowing the GPR to capture local curvature where the polynomial alone struggles to do so flexibly.

As shown in Figure~\ref{fig:Cont_fit_NIRSpec_both_epochs}, the GPR-produced continuum (black dashed line) smoothly follows the observed spectral shape in both quiescent and burst phases.
Subtracting this baseline yields robust optical depth spectra (bottom panel), which form the basis for our global ice analysis.
The uncertainty in the continuum was estimated directly from the predictive variance of the GPR model, and this continuum uncertainty was propagated together with the flux uncertainty to derive the uncertainty in optical depth spectra.
In addition to the continuum uncertainty, the presence of numerous narrow gaseous absorption and emission lines introduces localized increases in the optical depth uncertainty. 
These features are present in both epochs, but are more pronounced in the quiescent phase, particularly for CO and H$_2$O, where deeper line absorption reduces the local flux level and amplifies the relative uncertainty (Paper II).
A weak scattering feature at \microns{4.2} on the short-wavelength side of the CO$_2$ ice feature is interpreted as a signature of grain growth, consistent with the accompanying red wing structure \citep{Dartois2024}.
A detailed analysis of these grain growth signatures and their implications will be explored in a forthcoming study.

\subsection{MIRI MRS}
Accurate determination of the continuum baseline for MIRI MRS spectra is critical because of the extensive overlap between molecular ice and broad silicate absorption.
Mid-infrared ice studies have employed a sequential silicate subtraction approach, which fits a smooth continuum baseline first, then subtracts pre-determined models of the silicate absorption profile \citep{Dorschner1995,Boogert2011,Poteet2015} or the silicate template spectrum \citep{Kemper2004}, to isolate molecular ice features \citep{Boogert2008,Bottinelli2010}.
Recent JWST-based ice studies \citep{McClure2023,chen2024,rocha2024,vanDishoeck2025} similarly used such a method to subtract silicate absorption from MIRI spectra.
Our approach integrates continuum determination and silicate subtraction into a single, unified fitting process.
We simultaneously fit an initial fourth-order polynomial baseline along with a synthetic silicate absorption profile constructed from laboratory opacities of amorphous pyroxene (Mg$_{0.7}$Fe$_{0.3}$SiO$_3$) and olivine (MgFeSiO$_4$), consistent with compositions widely applied in previous mid-infrared ice analyses \citep{Boogert2011,Poteet2015,McClure2023,rocha2024}.
Specifically, the silicate models were synthesized using the $optool$ \citep{Dominik2021}, incorporating grain opacities with a size distribution from 0.1 to 1 $\mu$m following a power-law index of 3.5. The resulting dust composition consists of 87\% silicates and 13\% carbonaceous dust.
We optimized the silicate absorption fit by incorporating the \microns{9} ice absorption into the spectrum and corrected the continuum.
Ammonia-based ice components, such as pure ammonia (NH$_3$) and an H$_2$O-rich NH$_3$ mixture, have been considered the dominant ice species in the \microns{9} region \citep{Gibb2004,Bottinelli2010}.

\begin{figure}[htp!]
    \centering
    \includegraphics[width=1.0\textwidth]{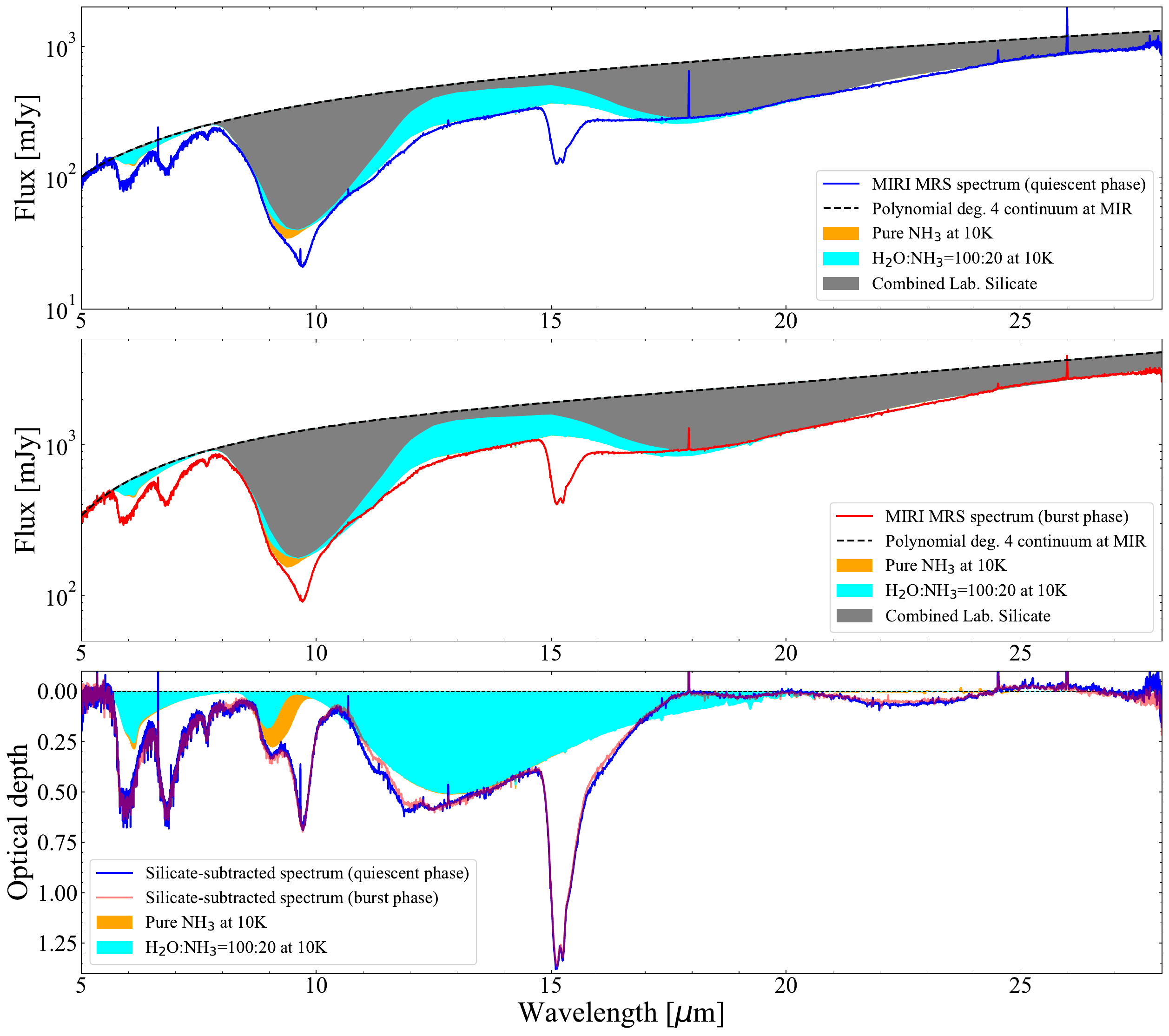}
    \caption{MIRI MRS spectrum of EC 53 observed at quiescent phase (top) and burst phase (middle). The continuum curve fitted by a fourth-order polynomial is adjusted using the combined silicate absorption profile derived from laboratory opacities of amorphous pyroxene (Mg$_{0.7}$Fe$_{0.3}$SiO$_3$) and olivine (MgFeSiO$_4$) to reproduce the absorption features near \microns{10} and \microns{18}. The NH$_3$-based ice features (the cyan- and orange-shaded regions for H$_2$O-rich and pure components, respectively) are used to optimize the fitted continuum covering the \microns{9} absorption as well as the broad absorption region at the H$_2$O libration mode. (Bottom) Converted optical depth spectrum for the ice absorption features of both phases. }
    \label{fig:miri_continuum}
\end{figure}

Figure~\ref{fig:miri_continuum} explicitly illustrates our simultaneous continuum and silicate fitting procedure for both the quiescent (top panel) and burst phases (middle panel) of EC 53. It demonstrates how the polynomial continuum (dashed line) directly combines with the fitted silicate absorption (gray-shaded region) to reproduce the observed spectrum effectively. Notably, the prominent ammonia-based ice feature around \microns{9}, which is represented by laboratory spectra of pure NH$_3$ ice (orange) and H$_2$O-mixed NH$_3$ ice (cyan), is carefully anchored to avoid artificial distortion of the continuum baseline.
The bottom panel of Figure~\ref{fig:miri_continuum} shows the resulting silicate-subtracted optical depth spectra, clearly isolating the ice absorption features. The residual spectra exhibit robust, well-defined ice absorption bands, suitable for subsequent quantitative analysis and global ice decomposition. 
We estimated the continuum uncertainty using a bootstrap analysis of the polynomial fitting procedure, and propagated this uncertainty together with the flux uncertainty to derive the ice absorption uncertainty of the optical depth spectrum. 
This approach provides a quantitative estimate of the uncertainty associated with continuum placement in the presence of broad silicate absorption and overlapping ice features.
By fitting continuum and silicate absorption simultaneously, we accurately isolate the ice absorption features behind broad silicate absorption, thereby enhancing the reliability of derived ice abundances compared to traditional sequential approaches.

\begin{figure}[hp!]
    \centering
    \includegraphics[height=1.05\textwidth]{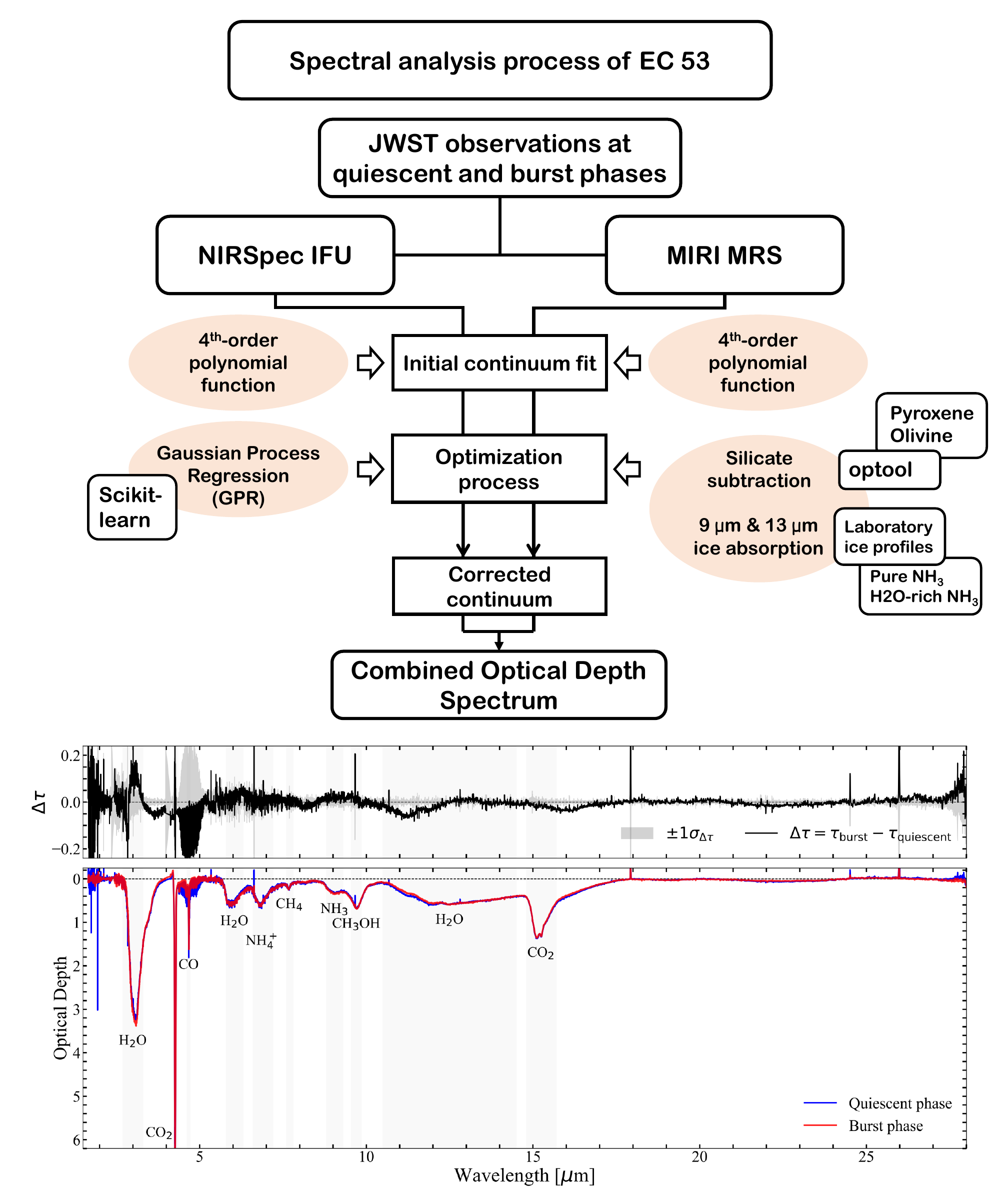}\
    \caption{Flowchart summarizing the continuum determination and spectral decomposition procedure applied to the JWST NIRSpec IFU and MIRI MRS observations of EC~53 in both quiescent and burst phases. These steps are combined within an optimization framework to derive the final corrected continuum and the combined optical depth spectrum. The middle panel shows the difference in optical depth spectra between the burst and quiescent phases, defined as $\Delta\tau = \tau_{\mathrm{burst}} - \tau_{\mathrm{quiescent}}$, with the gray shaded region indicating the propagated 1$\sigma$ uncertainty in the difference spectrum. Bottom panel plots the resulting optical depth of the quiescent and burst phases over the full wavelength range. The labeled shaded bands mark the major ice absorption features used in the comparison.}
    \label{fig:flowchart_cont}
\end{figure}

\section{Results \label{sec:results}}
The continuum determination process for both NIRSpec and MIRI spectra of EC 53 is summarized in Figure~\ref{fig:flowchart_cont}, which illustrates the step-by-step procedure for deriving the combined optical depth spectrum after silicate subtraction.
Applying this procedure independently to the quiescent-phase and burst-phase spectra yields optical depth spectra of the ice absorption features with uncertainties directly propagated from the flux and continuum estimates.

The resulting optical depth difference (middle panel in Figure~\ref{fig:flowchart_cont}), defined as $\Delta\tau = \tau_{\mathrm{burst}} - \tau_{\mathrm{quiescent}}$, generally remains close to zero across the major ice absorption bands in the mid-infrared region, including H$_2$O, CO$_2$, CO, NH$_3$, and CH$_3$OH, and is largely consistent with the propagated 1$\sigma$ uncertainty.
In the near-infrared region, however, localized deviations beyond the 1$\sigma$ level become more evident due to the presence of numerous gaseous absorption and emission lines. These features are present in both epochs, but more prominent in the quiescent phase, particularly for CO and H$_2$O (Paper II), where deeper line absorption leads to increased uncertainty in the derived optical depth.
As a result, the localized deviations in $\Delta\tau$ are dominated by noise and line contamination rather than intrinsic differences in the ice absorption bands.
The quiescent-phase spectrum also exhibits a relatively low signal-to-noise ratio, which further increases the scatter in the derived optical depth and its uncertainty.

Given these limitations, the ice absorption features in the quiescent and burst phases are generally consistent within the observational uncertainties.
Nevertheless, the burst-phase spectrum, with its higher signal-to-noise ratio and reduced line contamination, provides a more robust basis for quantitative analysis. Therefore, in the following sections, we primarily base our discussion on the burst-phase results, while using the quiescent-phase spectrum as a consistency check.

\subsection{Ice Features \label{sec:results_ices}}
In the bottom panels of Figures~\ref{fig:Cont_fit_NIRSpec_both_epochs} and~\ref{fig:miri_continuum}, the absorption features of various ice species appear in the spectrum corresponding to their vibration modes.
The spectrum exhibits distinct and broad absorption features corresponding to the stretching and libration modes of H$_2$O ice in the wavelength ranges of 2.7–\microns{3.4} and 10–\microns{20}, respectively \citep{DHendecourt1986,Bouilloud2015}. 
The bending mode of H$_2$O ice at \microns{6.0} is blended with organic ice components like formic acid (HCOOH), formaldehyde (H$_2$CO), and acetaldehyde (CH$_3$CHO), resulting in complex absorption features \citep{Boogert2008}. 
In addition, laboratory studies by \citet{Slavicinska2023} have shown that formamide (NH$_2$CHO ) ice contributes not only to the broad complex around \microns{6} but also exhibits distinct absorption features near 7.2 and \microns{7.5}. 

The bending mode of CO$_2$ ice near \microns{15} exhibits a distinctive double-peaked absorption feature, which primarily originates from its pure component \citep{2008pontoppidan,Kim2012}, accompanied by a shoulder extending toward longer wavelengths. 
Another CO$_2$ ice band appears as a sharp absorption feature centered at \microns{4.27} due to the stretching mode.
CO ice, on the other hand, manifests only in its stretching mode, producing a narrow absorption feature at \microns{4.67}. 
Additionally, the absorption feature associated with the OCN$^-$ ion at \microns{4.62} is clearly distinguishable due to the high spectral resolution of the NIRSpec spectrum.

The spectral region between 6.5 and \microns{8} observed in the MIRI MRS spectra of EC 53 provides rich diagnostics of various molecular ice species in protostellar environments.
A prominent absorption feature centered around \microns{6.85} clearly appears in this range, primarily attributed to ammonium ions (NH$_4^+$), formed by acid-base reactions involving NH$_3$ and acidic ice species such as HCOOH and H$_2$CO \citep{Schutte2003,Oberg2011,Rocha2025,Slavicinska2025}.
Previous laboratory experiments consistently suggested ammonium-based ice absorption near \microns{6.85} as a common feature toward embedded protostars \citep{Novozamsky2001,Schutte2003,Slavicinska2025}.
Additionally, subtle absorption bands in this region could be attributed to H$_2$CO \citep{Raunier2004} and the methanol-based ice species \citep[i.e. CH$_3$OH, CH$_3$CHO, and CH$_3$CH$_2$OH;][]{Oberg2011,Terwisscha2018,Yang2022,McClure2023}, which have absorption features near this range and potentially contribute to the complexity of the ice mixture observed in protostellar envelopes \citep{Oberg2011}.
The presence and relative abundance of these minor species have important implications for the ice chemistry and physical conditions within the protostellar envelope \citep{Boogert2015}.

The spectral interval from approximately 7.0 to \microns{7.5} contains additional subtle but complex absorption features. These absorption bands potentially arise from more intricate molecular species, including small amounts of CH$_3$CHO, ethanol (CH$_3$CH$_2$OH), and other COMs known to exhibit weak absorptions within this wavelength range \citep{Schutte1999,Oberg2011,Terwisscha2018}.
Because the absorption features in this region are weak and contaminated by gaseous absorption lines, it is challenging to determine the detailed ice composition.
The observed spectra exhibit distinct absorption around \microns{7.7}, characteristic of methane (CH$_4$). The detection and analysis of the CH$_4$ ice band at \microns{7.67} has proven significant in tracing volatile carbon-bearing ice reservoirs and assessing ice chemistry under varying physical conditions in star-forming regions \citep[e.g.,][]{Boogert1997, Oberg2008, Oberg2011, Boogert2015}. 
The strength and profile of this CH$_4$ ice absorption remain consistent between the quiescent and burst phases, indicating that the thermal alteration during the burst was insufficient to desorb the solid CH$_4$ component.

In the 8-10 $\mu$m range, where most of the absorption by the silicate components has been subtracted from the mid-infrared spectrum, broad absorption features are detected, primarily attributed to the ice components of the NH$_{3}$ umbrella mode and the CH$_{3}$OH stretching mode \citep{Bottinelli2010}.
These molecules also exhibit stretching and combination bands in the near-infrared \citep{Bouilloud2015}, with generally weak features appearing especially within the 2.9 and 3.2-\microns{3.8} regions, respectively.
However, identifying these bands is particularly challenging due to significant overlaps with the broad and intense H$_2$O ice absorption around \microns{3}, as well as additional contributions from other ice species such as ammonia hydrate \citep[NH$_3$.H$_2$O;][]{Moore2007}, CH$_4$ \citep{Boogert2004}, H$_2$CO \citep{Schutte1993}, and other COMs \citep{Terwisscha2018}.
Moreover, the region beyond \microns{3.2} is also associated with grain growth effects \citep{Smith1989,Dartois2024}, which overlap with these complex ice absorption bands, making it difficult to isolate the intrinsic spectral signatures of individual species.

\subsection{Global ice fitting process \label{sec:results_global}}
Once the optical depth spectrum was derived from the silicate-subtracted spectra, we conducted a global analysis of the ice composition.
We used laboratory absorbance profiles to decompose the observed absorption features into component spectra of individual ice species.
This decomposition employed scale coefficients determined through a Bayesian Markov Chain Monte Carlo (MCMC) fitting approach, implemented with the {\tt emcee} Python package \citep{Foreman-Mackey2013}, which ensured robust, physically meaningful correlations among ice components (see Appendix~\ref{sec:appendix_B} for detailed procedures).

Through this fitting methodology, we aimed to constrain both the chemical composition and thermal history of the icy mantles within the protostellar envelope of EC~53.
Given the complexity and overlap of ice absorption bands, it was essential to systematically isolate the dominant spectral contributions of major ice species within defined wavelength intervals.
This approach allowed us to account for the diverse vibrational modes of each species and facilitated interpretation of spectral signatures across the near- to mid-infrared range \citep{Kim2025}.
The laboratory ice profiles utilized as references for this analysis are listed in Table~\ref{tbl:lab_ref}.
Prior to fitting, all laboratory ice spectra underwent careful baseline corrections and were interpolated to match the resolution of the EC~53 optical depth spectrum, as detailed in Appendix~\ref{sec:appendix_A}.

\begin{figure}[hp!]
    \centering
    \includegraphics[scale=0.25]{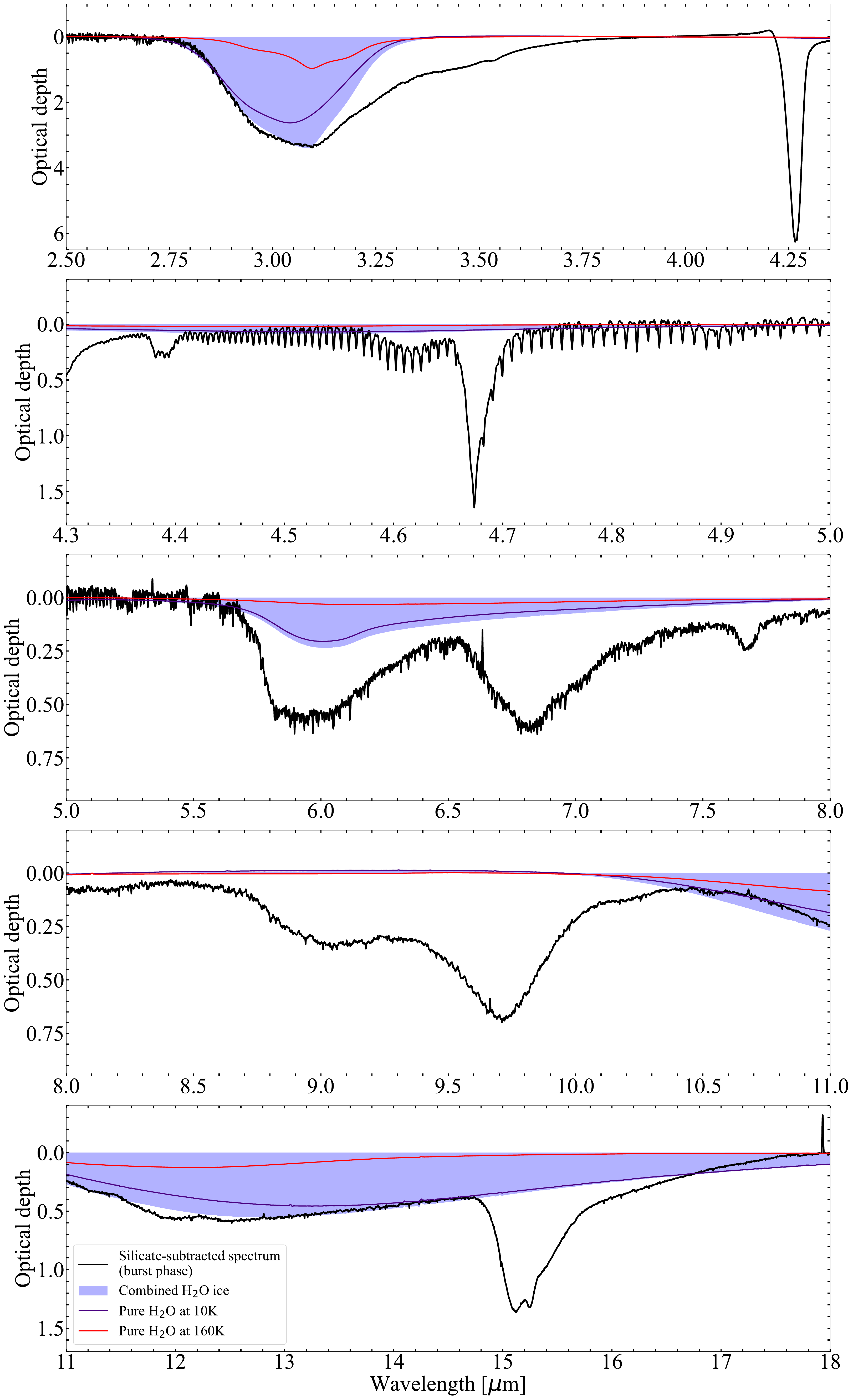}
    \caption{Ice composition of the pure H$_2$O with a combination of two temperatures (10~K and 160~K) at the entire spectral range. From top to bottom, panels show the NIRSpec and MIRI spectral coverage, highlighting the dominant absorption bands. }
    \label{fig:Ice_composition_water}
\end{figure}

The global fitting process began with H$_2$O ice, the most abundant ice species typically found in environments before and after the onset of star formation \citep{Boogert2008,Boogert2011}.
As illustrated in Figure~\ref{fig:Ice_composition_water}, the H$_2$O ice absorption was modeled by combining amorphous and crystalline ice components at temperatures of 10~K and 160~K, respectively. This dual-component approach effectively reproduced the shifted absorption peak near \microns{3.1} and the broad libration mode extending beyond \microns{10}, highlighting the importance of the crystalline component of the H$_2$O ice \citep{Hagen1983}.

After isolating the contribution of H$_2$O ice, several distinct ice absorption features became clearly apparent.
To systematically disentangle the contributions of various ice species throughout the spectral range, we sequentially analyzed absorption features from other major ice species, including CO$_2$, CH$_3$OH, NH$_3$, and CO.
Guided by previous research \citep{2008pontoppidan,Bottinelli2010,Aikawa2012,Noble2013,Kim2022,Kim2025,McClure2023,Rocha2025}, we primarily employed H$_2$O-rich mixtures, as pure ice profiles alone were insufficient to reproduce certain spectral features (see Figure~\ref{fig_appx:Ice_Lab_optim}).
However, because these mixed ice profiles inherently include contributions from H$_2$O ice, the intrinsic H$_2$O absorption component was subtracted after the pure H$_2$O ice contribution had been independently accounted for.
The detailed procedure for subtracting the intrinsic H$_2$O ice contribution from the mixed profiles is described in Appendix~\ref{sec:appendix_A} and illustrated in Figure~\ref{fig_appx:Ice_Lab_H2O_subt}.

We first manually applied the CH$_3$OH- and CO$_2$-based ice profiles to simultaneously fit the absorption bands at \microns{15}, \microns{9.7}, and \microns{4.27}.
The \microns{15} absorption band corresponds to the bending mode of CO$_2$ ice, superimposed on the broad H$_2$O libration absorption.
To accurately reproduce this region, we fitted a combination of laboratory-experimented CO$_2$ ice profiles, including pure CO$_2$ ice, as well as mixtures with H$_2$O, CO, and CH$_3$OH. 
These profiles were selected to best match the characteristic double-peaked structure near \microns{15.2} and its extended red wing, both of which are sensitive indicators of thermal processing and the ice matrix environment \citep{2008pontoppidan}.
In particular, the same combination of the CO$_2$ ice profile consistently accounted for the asymmetric stretching mode near \microns{4.27}, as well as the isotropic $^{13}$CO$_{2}$ feature at \microns{4.38} (Figure~\ref{fig_appx:Global_fit_CO2}). 
This consistency in the bending and stretching modes reinforces the robustness of our chosen CO$_2$ mixtures, supporting the reliability of the global decomposition analysis (Figure~\ref{fig_appx:Ice_fit_step2}).

Subsequently, for the CH$_3$OH ice composition, we incorporated pure and CO-mixed CH$_3$OH (CO:CH$_3$OH=1:1) profiles to reproduce the absorption at \microns{9.7} based on the intensity of the fitted CH$_3$OH mixture profile (H$_2$O:CO$_2$:CH$_3$OH=1:1:1) that was applied to the analysis of the CO$_2$ bending mode (see the detailed process of the ice composition in Appendix~\ref{subsec:appendix_B3} and the result in Figure~\ref{fig_appx:Ice_fit_step3}).
The initial ambiguity between the pure and CO-mixed CH$_3$OH profiles was resolved by fitting the CO absorption at \microns{4.67}, confirming the significant contribution of the CO-mixed CH$_3$OH ice profile (see the detailed ice composition process in Appendix~\ref{subsec:appendix_B4} and Figure~\ref{fig_appx:Ice_fit_step4}).
Figure~\ref{fig_appx:Ice_fit_step5} summarizes the resulting global ice composition fits for these bands.

The NH$_3$-based ice absorption within 8.5-\microns{10} was also refined considering additional tentative contributions from pure CH$_3$CH$_2$OH, in order to avoid overestimating pure NH$_3$ absorption (Figure~\ref{fig_appx:Ice_fit_step3}).
However, we caution that this fit is tentative, as the absorption feature near \microns{9.17}, which is characteristic of CH$_3$CH$_2$OH, does not provide conclusive support for a distinct CH$_3$CH$_2$OH identification. 
Based on the absorption characteristics near \microns{9}, we minimized contributions from CH$_3$CHO ice, but weak CH$_3$CHO ice absorptions might still contribute at other wavelengths such as \microns{5.8} and \microns{7.1} \citep{Terwisscha2018}.

In the adjacent \microns{4.67} region, the CO stretching mode appears as a distinct and narrow absorption feature, often accompanied by the OCN$^-$ absorption feature at \microns{4.62}, a known marker of energetic or thermal processing in nitrogen-bearing ices \citep{vanBroekhuizen2005}. 
Laboratory OCN$^-$ ice profiles at 12~K and 80~K were considered, but only the low-temperature (12~K) profile (light green-dashed line in Figure~\ref{fig_appx:Ice_fit_step4}) significantly reproduced the observed absorption feature at \microns{4.62}.  

To accurately fit the complex spectral region between 6 and \microns{8}, we considered overlapping contributions from multiple ice components, including H$_2$O-rich CH$_4$ and CH$_3$COOH mixtures, pure H$_2$CO, and the tentative inclusion of NH$_2$CHO , as described in Appendix~\ref{subsec:appendix_B6}.
Specifically, H$_2$O-rich CH$_4$ and CH$_3$COOH mixtures reproduced the \microns{7.7} band.
To constrain absorption near \microns{5.8} and \microns{6.7}, we included a pure H$_2$CO ice profile, reflecting its molecular bending modes.
A mixture of CO and HCOOH in the H$_2$O-rich ice profile (H$_2$O:CO:HCOOH = 62:30:8) helped reproduce weak absorption near 8.0–\microns{8.5} and contributed significantly near \microns{6}, particularly the blue wing of this absorption band, which overlaps with several organic species.
To account for weak absorption features at \microns{7.20} and \microns{7.53}, and to improve the fit at the shorter wavelength side of the \microns{6} band, we tentatively included a pure NH$_2$CHO  ice profile, although the vibrational signatures remain inconclusive. 
The OCN$^-$ ice profile at 12~K, previously applied to the \microns{4.62} feature, contributed slightly to absorption at \microns{7.62}.
Additionally, we fitted NH$_4^+$ ice to reproduce the broad absorption centered at \microns{6.85}, consistent with previous identifications in processed interstellar ices \citep{Schutte2003,Oberg2011,Rocha2025}.
The resultant fit showed strong consistency throughout the entire 6--\microns{8} spectral range (Figure~\ref{fig_appx:Ice_fit_step6}), indicating that the dominant ice absorption features in this complex region are well reproduced by the adopted combination of laboratory ice profiles.
Finally, we addressed residual discrepancies identified around the \microns{3} absorption region by conducting a refined fit specifically targeting the previously derived H$_2$O ice intensities (Appendix~\ref{subsec:appendix_B7}).

\begin{figure}[hp!]
    \centering
    \includegraphics[scale=0.25]{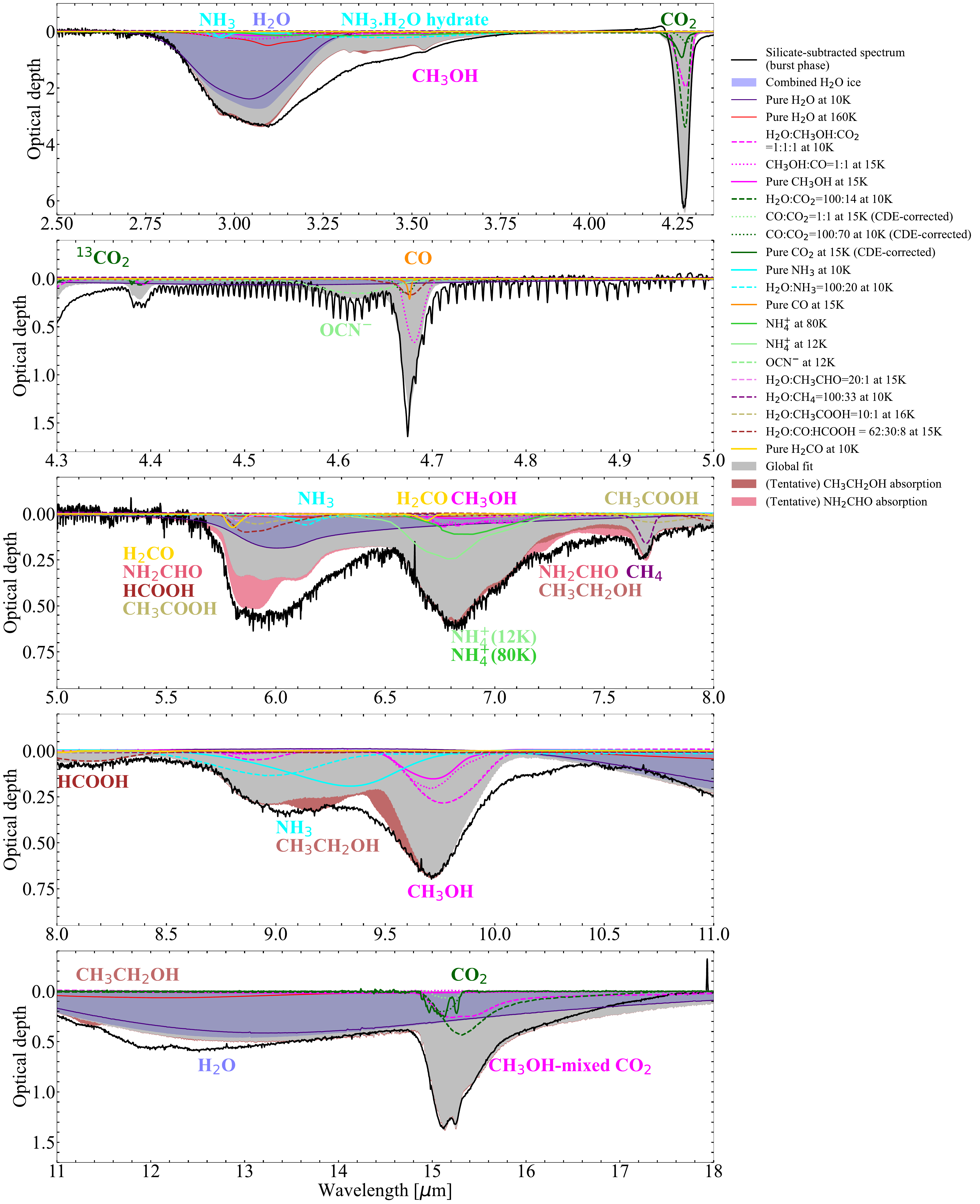}\
    \caption{Final result of the global ice fitting for EC~53. The silicate-subtracted spectrum obtained during the burst phase is shown in black. The gray shaded region represents the total modeled ice absorption constructed by summing the absorption profiles of all identified ice components on top of the combined H$_2$O ice baseline. The combined H$_2$O component is shown as a blue shaded region, while individual contributions from CO$_2$ (dark green lines), CO (dark orange lines), CH$_3$OH (magenta lines), NH$_3$ (cyan lines), and various complex organic molecules are overlaid. Tentative absorption features attributed to CH$_3$CH$_2$OH and NH$_2$CHO  are indicated by light-brown and pink shaded regions, respectively.}

    \label{fig:Ice_composition_all}
\end{figure}

Based on the optimized global fit (Figure~\ref{fig:Ice_composition_all}), the column densities (\textit{N}) for each ice component were calculated by integrating the optical depths of the best-fit laboratory profiles listed in Table~\ref{tbl:lab_ref}, using Equation~\ref{eq:3} in Appendix~\ref{sec:appendix_B}.
The resulting column densities, derived by applying the global ice fitting procedure to both phase spectra, are summarized in Table~\ref{tbl:col_den1_rev} and Table~\ref{tbl:col_den2_rev}.
The quoted uncertainties represent the statistical uncertainties of the MCMC fit under a fixed continuum and model assumption.
However, additional systematic uncertainties arise from continuum placement and degeneracies among different ice components, which are not fully captured in the formal MCMC uncertainties.
Taking these effects into account, the differences in the column densities between the two epochs are smaller than or comparable to the combined uncertainties. 
In particular, the total column densities of the major ice species differ only at the few-percent level between the two epochs, indicating that the inferred ice compositions are statistically consistent.

Our final refinement significantly reduced residual absorption features and yielded an optimized global ice composition that closely matched the observed optical depth spectrum across most of the wavelength range.
Nevertheless, several systematic residuals remain.
In the near-infrared range, residual absorption persists between \microns{3.2} and \microns{3.5}, a feature that has been widely interpreted as a signature of grain growth arising from changes in dust opacity and ice mantle structure \citep{Dartois2024}.
Residual absorption is also present in the red wing of the CO$_2$ stretching mode at \microns{4.27}, which is similarly linked to the grain growth effect.
We further note that the broad absorption attributed to organic refractory material, which has been proposed as a residual component following energetic processing of interstellar ices in protostellar envelopes \citep{Greenberg1995,Gibb2002,Boogert2008,rocha2024}, might contribute to the remaining absorption in the 6.0--\microns{6.5} region, where the global ice analysis does not yield a fully satisfactory fit.
Finally, Appendix~\ref{sec:appendix_C} discusses the possible contribution of crystalline silicate absorption to the residual features in the 9--\microns{10} region, which likewise remain imperfectly reproduced by the global ice fit.

\begin{deluxetable*}{cccccc}
\centering
\tablecaption{laboratory ice profiles for the global fitting process}
\label{tbl:lab_ref}
\tablehead{\colhead{Species} & \colhead{\textit{A}\tablenotemark{a} (cm mole$^{-1}$)} & \colhead{$\lambda_{peak}$ ($\mu$m)} & \colhead{Temperature (K)} & \colhead{References} & } 
\startdata
\multicolumn{6}{c}{Complex ice mixtures} \\
\hline
H$_2$O:CH$_3$COOH = 10:1                       & 4.57E-17 & 7.77 & 16 & \citet{Hudson1999}\tablenotemark{b} & \\
H$_2$O:CH$_3$CHO = 20:1                    & 3.0E-17 & 5.81  & 15 & \citet{Terwisscha2018} & \\
H$_2$O:CO:HCOOH = 62:30:8          & 1.1E-17 & 4.67 (CO)  & 15 & \citet{Bisschop2007} & \\
                                               & 2.9E-17 & 8.15 (HCOOH)  & & & \\
\hline
\multicolumn{6}{c}{Pure complex ice (Tentative components)} \\
\hline
CH$_3$CH$_2$OH                                 & 7.35E-18 & 9.17  & 15 & \citet{Terwisscha2018} & \\
NH$_2$CHO                                      & 0.9E-17 & 7.20  & 15 & \citet{Slavicinska2023} & \\
\hline
\multicolumn{6}{c}{Simple ice mixtures} \\
\hline
CO:CH$_3$OH = 1:1                              & 1.1E-17 & 4.68 (CO) & 15 & \citet{Fraser2004} & \\
                                               & 1.8E-17 & 9.70 (CH$_3$OH) &  &  & \\
CO:CO$_2$ = 1:1\tablenotemark{c}                                & 1.1E-17 & 4.66 (CO) & 15 & \citet{vanBroekhuizen2006} & \\
                                               & 1.2E-17 & 15.20 (CO$_2$) & &  & \\
CO:CO$_2$ = 100:70\tablenotemark{c}                             & 1.1E-17 & 4.66 (CO) & 10 & \citet{Ehrenfreund1997} & \\
                                               & 1.2E-17 & 15.20 (CO$_2$) &  &  & \\
H$_2$O:CO$_2$ = 100:14                         & 1.2E-17 & 15.30 & 10 & \citet{Ehrenfreund1997} & \\        
H$_2$O:CH$_4$ = 100:33                         & 8.4E-18 & 7.69  & 10 & \citet{Ehrenfreund1997} & \\      
H$_2$O:NH$_3$ = 100:20                         & 1.63E-17 & 8.96  & 10 & \citet{Ehrenfreund1997}& \\ 
H$_2$O:CO = 100:20                             & 1.1E-17 & 4.67  & 10 & \citet{Ehrenfreund1997} & \\ 
H$_2$O:CH$_3$OH:CO$_2$ = 1:1:1           & 1.8E-17 & 9.76 (CH$_3$OH)  & 10 & \citet{Ehrenfreund1999} & \\
                                               & 1.2E-17 & 15.20, 15.38 (CO$_2$) &  &  & \\
\hline
\multicolumn{6}{c}{Pure simple ice} \\
\hline
H$_2$O                                         & 2.2E-16 & 3.05, 3.10  & 10, 160 & \citet{Gerakines1996} & \\
                                               & 3.2E-17 & 13.00, 12.2  &  & & \\
NH$_4^+$                                       & 4.4E-17 & 6.79, 6.87 & 12, 80 & \citet{Novozamsky2001} & \\
OCN$^-$                                        & 1.3E-16 & 4.62 & 12 & \citet{Novozamsky2001} & \\
H$_2$CO                                        & 1.5E-18 & 8.02 & 10 & \citet{Gerakines1996} & \\
CO$_2$\tablenotemark{c}                                         & 1.2E-17 & 15.12, 15.26 & 15 & \citet{vanBroekhuizen2006} & \\
CO                                             & 1.1E-17 & 4.67 & 15 & \citet{vanBroekhuizen2006} & \\
NH$_3$                                         & 1.63E-17 & 9.34 & 10 & \citet{Gerakines1996} & \\
\enddata
\tablecomments{Data taken from the Leiden Ice Database for Astrochemistry \citep[LIDA;][]{Rocha2022}}
\tablenotetext{a}{Each band strength of the ice component at the peak position of the designated vibrational mode is referred to as follows, \citet{Jin2022}: CO$_2$, CO, and CH$_3$CHO; \citet{Hudson2017}: CH$_3$CH$_2$OH; \citet{rocha2024}: CH$_3$COOH; \citet{Bouilloud2015}: H$_2$O, NH$_3$, CH$_4$, CH$_3$OH, HCOOH, and H$_2$CO; \citet{Schutte2003}: NH$_4^+$; \citet{Slavicinska2023}: NH$_2$CHO }
\tablenotetext{b}{NASA Goddard Cosmic Ice Laboratory}
\tablenotetext{c}{Corrected by the grain shape model using the continuous distribution of ellipsoids \citep[CDE;][]{2008pontoppidan}.}
\end{deluxetable*}

\begin{deluxetable*}{llcc}
\tablecaption{Ice column densities of major ice species\label{tbl:col_den1_rev}}
\tablewidth{0pt}
\tablehead{
\colhead{Species} & \colhead{Composition} & \colhead{Temperature (K)} & \colhead{$N$ ($10^{17}$ molecules cm$^{-2}$)}
}
\startdata
H$_2$O
  & Pure (amorphous)         & 10  & 37.85 $\pm$ 0.05 (36.28 $\pm$ 0.08)\\
  & Pure (crystalline)       & 160 & 4.46 $\pm$ 0.05 (4.34 $\pm$ 0.10)\\
  & Total                    &      & 42.31 (40.62)\\
  &                          &      &        \\
CO$_2$
  & Pure                     & 15  & 2.60 $\pm$ 0.03 (2.40 $\pm$ 0.05)\\
  & CO-mixed (1:1)           & 15  & 0.70 $\pm$ 0.001 (0.70 $\pm$ 0.003)\\
  & CO-mixed (100:70)        & 10  & 2.27 $\pm$ 0.004 (2.34 $\pm$ 0.007)\\
  & CH$_3$OH-mixed           & 10  & 6.07 $\pm$ 0.25 (5.92 $\pm$ 0.30)\\
  & H$_2$O-rich              & 10  & 9.33 $\pm$ 0.21 (9.93 $\pm$ 0.24)\\
  & Total                    &      & 20.97 (21.29) \\
  &                          &      &        \\
CO
  & Pure                     & 15  & 0.34 $\pm$ 0.03 (0.38 $\pm$ 0.03)\\
  & CO$_2$-mixed (1:1)       & 15  & 0.41 $\pm$ 0.001 (0.41 $\pm$ 0.001)\\
  & CO$_2$-mixed (70:100)    & 10  & 3.06 $\pm$ 0.04 (3.17 $\pm$ 0.05)\\
  & CH$_3$OH-mixed           & 15  & 5.34 $\pm$ 0.04 (5.30 $\pm$ 0.05)\\
  & H$_2$O-rich              & 15  & 2.10 $\pm$ 0.04 (2.97 $\pm$ 0.05)\\
  & Total                    &      & 11.25 (12.23)\\
  &                          &      &        \\
NH$_3$
  & Pure                     & 10  & 8.64 $\pm$ 0.02 (8.64 $\pm$ 0.03)\\
  & H$_2$O-rich              & 10  & 6.15 $\pm$ 0.01 (5.20 $\pm$ 0.03)\\
  & Total                    &      & 14.79 (13.84) \\
  &                          &      &        \\
CH$_4$
  & H$_2$O-rich              & 10  & 2.81 $\pm$ 0.05 (2.71 $\pm$ 0.07)\\
\enddata
\tablecomments{All column densities are given in units of 10$^{17}$ molecules cm$^{-2}$. Values outside parentheses correspond to the burst-phase spectrum, while values in parentheses indicate the results derived from the quiescent-phase spectrum using the same global fitting procedure. For species represented by multiple fitted components, the total column density is listed in a separate row.}
\end{deluxetable*}

\begin{deluxetable*}{llcc}
\tablecaption{Ice column densities of COM and ion species\label{tbl:col_den2_rev}}
\tablewidth{0pt}
\tablehead{
\colhead{Species} & \colhead{Composition} & \colhead{Temperature (K)} & \colhead{$N$ ($10^{17}$ molecules cm$^{-2}$)}
}
\startdata
CH$_3$OH
  & Pure                     & 15 & 2.48 $\pm$ 0.02 (2.38 $\pm$ 0.02)\\
  & CO-mixed                 & 15 & 2.92 $\pm$ 0.006 (2.90 $\pm$ 0.008)\\
  & CO$_2$-mixed             & 10 & 5.39 $\pm$ 0.01 (5.26 $\pm$ 0.02)\\
  & Total                    &    & 10.79 (10.54) \\
  &                          &    &      \\
CH$_3$CHO\tablenotemark{a}
  & H$_2$O-rich              & 15 & $<0.41$ ($<0.41$)\\
  &                          &    &      \\
CH$_3$CH$_2$OH\tablenotemark{a}
  & Pure                     & 15 & $<2.13$ ($<1.88$)\\
  &                          &    &      \\
CH$_3$COOH\tablenotemark{a}
  & H$_2$O-rich              & 15 & $<0.49$ ($<0.36$)\\
  &                          &    &      \\
NH$_4^+$
  & Pure                     & 80 & 2.05 $\pm$ 0.05 (2.19 $\pm$ 0.09)\\
  & Pure                     & 12 & 4.76 $\pm$ 0.06 (4.55 $\pm$ 0.11)\\
  & Total                    &    & 6.81 (6.74)\\
  &                          &    &      \\
OCN$^-$
  & Pure                     & 12 & 0.52 $\pm$ 0.001 (0.73 $\pm$ 0.005)\\
  &                          &    &      \\
HCOOH\tablenotemark{a}
  & H$_2$O-rich              & 15 & $<1.19$ ($<1.68$)\\
  &                          &    &      \\
H$_2$CO
  & Pure                     & 10 & 0.99 $\pm$ 0.009 (0.98 $\pm$ 0.015) \\
  &                          &    &      \\
NH$_2$CHO\tablenotemark{a}
  & Pure                     & 15 & $<1.14$ ($<0.85$)\\
\enddata
\tablecomments{All column densities are given in units of 10$^{17}$ molecules cm$^{-2}$. Values outside parentheses correspond to the burst-phase spectrum, while values in parentheses indicate the results derived from the quiescent-phase spectrum using the same fitting procedure. Ice components whose absorption bands are present in laboratory spectra but not clearly distinguished in the observed ice composition are reported only as upper limits. For species represented by multiple fitted components, the total column density is listed in a separate row.}
\tablenotetext{a}{Tentative identification.}
\end{deluxetable*}

\clearpage

\section{Discussions\label{sec:discussions}}
\subsection{Implications for Episodic Accretion and Ice Chemistry \label{sec:discussion_1}}
Our JWST spectra of EC~53, obtained in both quiescent and burst phases, show negligible variations in the ice absorption profiles. The minimal observed changes indicate that the periodic accretion bursts in EC~53 have little or no measurable impact on the ice composition over the $\sim$0.5~yr interval probed by our observations. This result provides valuable insights into the physical and chemical processes governing episodic accretion environments. Specifically, it underscores the importance of both burst amplitude and duration in determining whether ice species experience significant sublimation and recondensation cycles.

In EC~53, the modest burst amplitude and short recurrence interval fundamentally limit the extent of chemical evolution in the ice reservoir during a single burst cycle. 
Because bursts of similar magnitude recur every $\sim$1.5 years, the ice sublimation front in the envelope changes only modestly from one burst cycle to the next. 
Moreover, the freeze-out timescale of sublimated molecules in the inner envelope is expected to be much longer than the burst interval \citep{Visser2015,Rab2017}. 
For typical densities in the inner envelope, $n_{\rm H}\sim10^{5}$--$10^{6}$~cm$^{-3}$, the freeze-out timescale for common volatile species, such as CO, is on the order of $\sim10^{3}$--$10^{4}$~yr \citep{Jlee2004,Vorobyov2013,Visser2015,Rab2017}.
This timescale exceeds the $\sim1.5$~yr burst interval \citep{Hodapp2012,YLee2020} by several orders of magnitude, preventing efficient recondensation of sublimated material between successive bursts.
As a result, even if partial sublimation occurs locally during the burst, the limited spatial extent and short duration of heating confine the desorption zone, yielding net changes below our observational sensitivity.

\begin{figure}[hp!]
    \centering
    \includegraphics[scale=0.4]{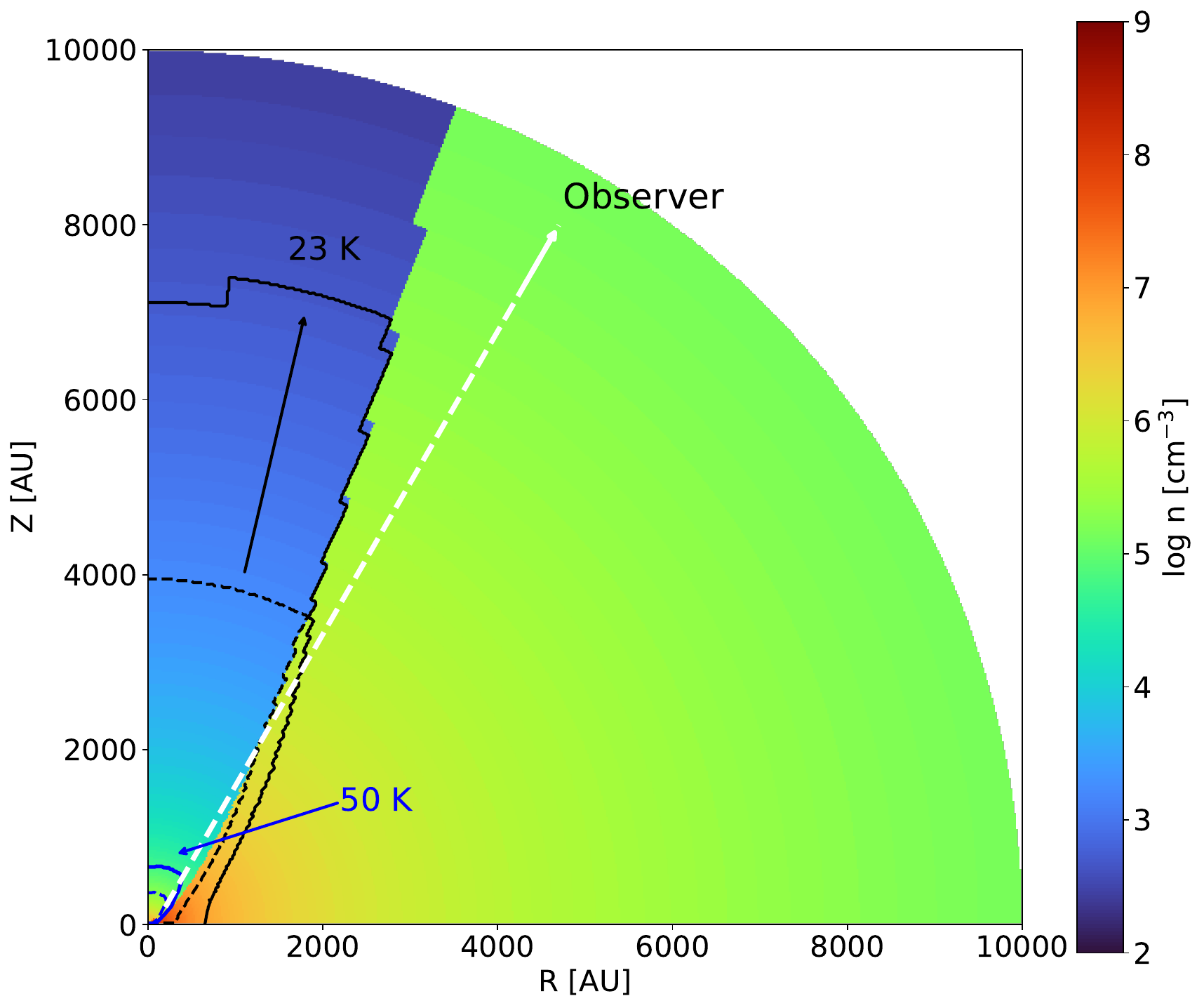}\
    \caption{Two-dimensional density and temperature structure of the EC~53 envelope derived from the radiative transfer model of \citep{Baek2020}. The background color map shows the gas number density, with the blue region indicating the outflow cavity and the green-to-orange region tracing the dense envelope. A representative line of sight toward the observer is indicated by a dashed arrow, illustrating the viewing geometry through the envelope. Temperature contours at $\sim$23~K and $\sim$50~K, representative of the sublimation temperatures of CO and CO$_2$ ices, respectively, are overlaid. Dashed contours denote the locations of these temperature thresholds in the quiescent phase, while solid contours show their outward expansion during the burst phase. The 100~K contour relevant for H$_2$O ice sublimation is confined to a very small region close to the protostar and is therefore not shown.}
    \label{fig:temperature_map}
\end{figure}

These conclusions are further supported by the spatial distribution of temperature and density predicted by radiative transfer modeling.
Figure~\ref{fig:temperature_map} shows the two-dimensional envelope structure from the radiative transfer model of \citet{Baek2020}, overlaid with temperature contours corresponding to $\sim$23~K and $\sim$50~K, representative of the approximate sublimation temperatures of CO and CO$_2$ ices, respectively.
In both the quiescent and burst phases, regions exceeding the CO sublimation temperature remain confined within $\lesssim$2000~au and are largely restricted to areas near the outflow cavity.
Although burst heating leads to a modest outward expansion of the CO and CO$_2$ sublimation fronts, as illustrated by the shift from dashed to solid contours in Figure~\ref{fig:temperature_map}, this expansion encompasses only a small fraction of the envelope volume.
Consequently, the difference in the sublimated ice volume between the two phases represents only a minor contribution to the total line-of-sight ice column probed by the observations.

Consistent with this picture, the radiative transfer modeling predicts that the internal luminosity of EC~53 increases by only a factor of $\sim$3.3 during the burst phase, substantially weaker than the luminosity enhancements observed in strongly eruptive systems such as EXor- or FUor-type outbursts \citep{Baek2020}. 
Such moderate heating shifts the sublimation fronts only slightly and affects a limited fraction of the line-of-sight column, further diluting any phase-dependent variation in integrated ice optical depths. 
For example, in the case of CO, the sublimation front shifts outward from \(\sim330\)~au to \(\sim660\)~au along the disk midplane. Even with this shift, the total ice mass decreases by only about \(2.1\%\) during the burst if CO ice in regions heated above \(T_{\rm sub}=23\,{\rm K}\) is assumed to sublimate into the gas phase.
Taking into account the source distance (436\,pc) and inclination (\(30^\circ\)), the line-of-sight CO-ice-bearing column density averaged over apertures of \(\sim300\text{--}600\)~au (roughly \(3\text{--}5\) times the JWST PSF \citep{Law2023} and comparable to the spatial scale of the extracted spectra) decreases by only \(\sim6.5\%\).
This modest change indicates that the outer, cooler envelope, which dominates the total column density, remains largely unaffected. Moreover, since we assume that CO ice within regions heated above the sublimation temperature is completely sublimated during the burst, the actual amount of newly sublimated ice is likely smaller; therefore, the fractional change derived here should be regarded as an upper limit. 
Consequently, the difference in the ice-sublimation region between the two phases occupies only a small fraction of the full line-of-sight path through the envelope, resulting in variations in the observable ice absorption that remain within the observation sensitivity.

The relatively low abundance of gaseous CH$_3$OH observed toward EC~53 \citep{SLee2020}, compared to strongly eruptive sources such as V883~Ori \citep{JLee2019}, reinforces the conclusion that episodic heating in EC~53 leaves only a minimal imprint on the ice reservoir.

Taken together, both the observational constraints and radiative transfer modeling indicate that short, moderate-amplitude bursts in EC~53 largely preserve the envelope ice reservoir, producing no measurable change in ice composition over the $\sim$0.5 yr interval probed.

\subsection{Cold-phase ice inventory and thermal evolution of ices toward EC~53}
Table~\ref{tbl:abundance} summarizes the ice abundances toward EC~53 relative to H$_2$O and compares them with values measured toward low-mass protostars \citep{Boogert2015} including JWST samples \citep{rocha2024,chen2024,Rocha2025}.
EC~53 stands out by its conspicuously high fractions of relative abundances of major ices, notably CO$_2$ (49\%), CH$_3$OH (25\%), NH$_3$ (34\%), and CO (26\%), exceeding the canonical ranges reported from past infrared surveys \citep[e.g.,][]{Boogert2008,2008pontoppidan,Bottinelli2010,Oberg2011}. 
Such an inventory is naturally explained if a large portion of the ice mantle accumulated during a cold prestellar phase.
In this regime ($T \lesssim$ 15~K), grain-surface hydrogenation of CO and N proceeds efficiently, building up CH$_3$OH and NH$_3$.
Consistent with this interpretation, our global ice composition finds that the column density of pure NH$_3$ is comparable to the H$_2$O-rich NH$_3$ mixture, indicating that a substantial portion of N-bearing ice likely formed in cold and shielded regions prior to the onset of protostellar heating.
Laboratory experiments and astrochemical models have shown that NH$_3$ forms efficiently via successive hydrogenation of atomic nitrogen on dust grains at temperatures below 15~K \citep{Fedoseev2015,Kakkenpara2025}. 
Similarly, CH$_3$OH forms through the stepwise hydrogenation of CO under cold conditions and is often enhanced in sources with prolonged cold phases.
The high abundance of CH$_3$OH in EC~53 further reinforces the interpretation of a chemically mature prestellar origin.

A useful context comes from the field star CK~2 behind the Serpens molecular cloud, which probes quiescent material along a line of sight.
Previous CK~2 measurements showed substantial abundances of CO and CO$_2$ ice, characteristic of cold, dense cloud conditions in the Serpens region \citep{Knez2005,Whittet2007}.
\citet{Perotti2020} also measured the high abundances of CO and CH$_3$OH relative to H$_2$O ice toward ten low-mass protostars in the Serpens SVS4 cluster, which is characterized by a cold cloud condition ($\sim$15~K).
These previous studies provide a plausible baseline for the possibility that EC~53 could inherit the chemically rich ice reservoir formed at low temperatures.

While many ice components likely formed under cold conditions, several independent diagnostics indicate thermal processing of the ices in EC~53, plausibly linked to intermittent luminosity bursts from episodic accretion, which reshaped the inherited cold-phase ice inventory within the Serpens cloud. 
The CO$_2$ bending mode near \microns{15.2} exhibits a prominent double-peaked absorption feature that is best reproduced with a CDE-corrected pure CO$_2$ ice profile (Figure~\ref{fig_appx:Global_fit_CO2}).
This characteristic arises from distillation and segregation within CO-mixed ices at $\sim$ 20$-$30~K, volatile CO desorbs from the CO$-$CO$_2$ mixtures, leaving behind segregated CO$_2$ ice \citep{2008pontoppidan,Oberg2009}.
The prominent peak of pure CO$_2$ ice absorption in the EC~53 spectrum suggests that its envelope has experienced thermal processing consistent with the warm-up episodes expected during accretion bursts.
Further support comes from the H$_2$O libration mode, which required modeling with both 10~K amorphous and 160~K crystalline ice profiles (Appendix~\ref{subsec:appendix_B7}).
This combination suggests partial crystallization of amorphous H$_2$O ice via thermal annealing, which is another indicator of heating events beyond cold prestellar conditions.

Additional processing tracers include the strong, broad absorption near \microns{6.8}, assigned to NH$_4^+$.
Our global fits favor a temperature-dependent redshift from $\sim$\microns{6.80} (12~K) to $\sim$\microns{6.85} (80~K), matching laboratory trends \citep{Keane2001,Schutte2003,Boogert2008}.
The comparable NH$_4^+$ abundance (16\%) to other low-mass protostellar samples in Table~\ref{tbl:abundance} and the clear detection of OCN$^-$ at \microns{4.62} together indicate acid-base and radical chemistry driven by thermal and energetic processing \citep{vanBroekhuizen2004,Oberg2011}. 

The tentative identification of minor COMs such as CH$_3$COOH ($\sim$1\%) and HCOOH ($\sim$2\%) is consistent with a mild warm-up that enables radical diffusion and recombination on/in the ice, superposed on a mantle already enriched by cold hydrogenation pathways. 
The coexistence of cold-phase products with thermally evolved components implies that the ice inventory toward EC~53 retains a layered chemical history. 
In this layered structure, the deeper, shielded strata preserve pristine cold-phase products (e.g, pure NH$_3$). In contrast, the outer layers record episodic heating and energetic processing, which together produce the exceptionally high and chemically diverse ice inventory observed toward EC~53.

\subsection{Comparison with other JWST ice studies and robustness of the EC~53 ice inventory}
The high ice abundances derived toward EC~53 raise the question of whether this source represents an extreme but physically meaningful case within the emerging JWST ice inventory, or whether these results could be influenced by methodological choices in continuum determination and silicate subtraction.
To place EC~53 in a broader observational context and to assess the robustness of our analysis, we compare our results with other JWST ice studies summarized in Table~\ref{tbl:abundance}, with particular emphasis on the well-studied protostellar source B1-c.

\begin{figure}[hp!]
    \centering
    \includegraphics[scale=0.4]{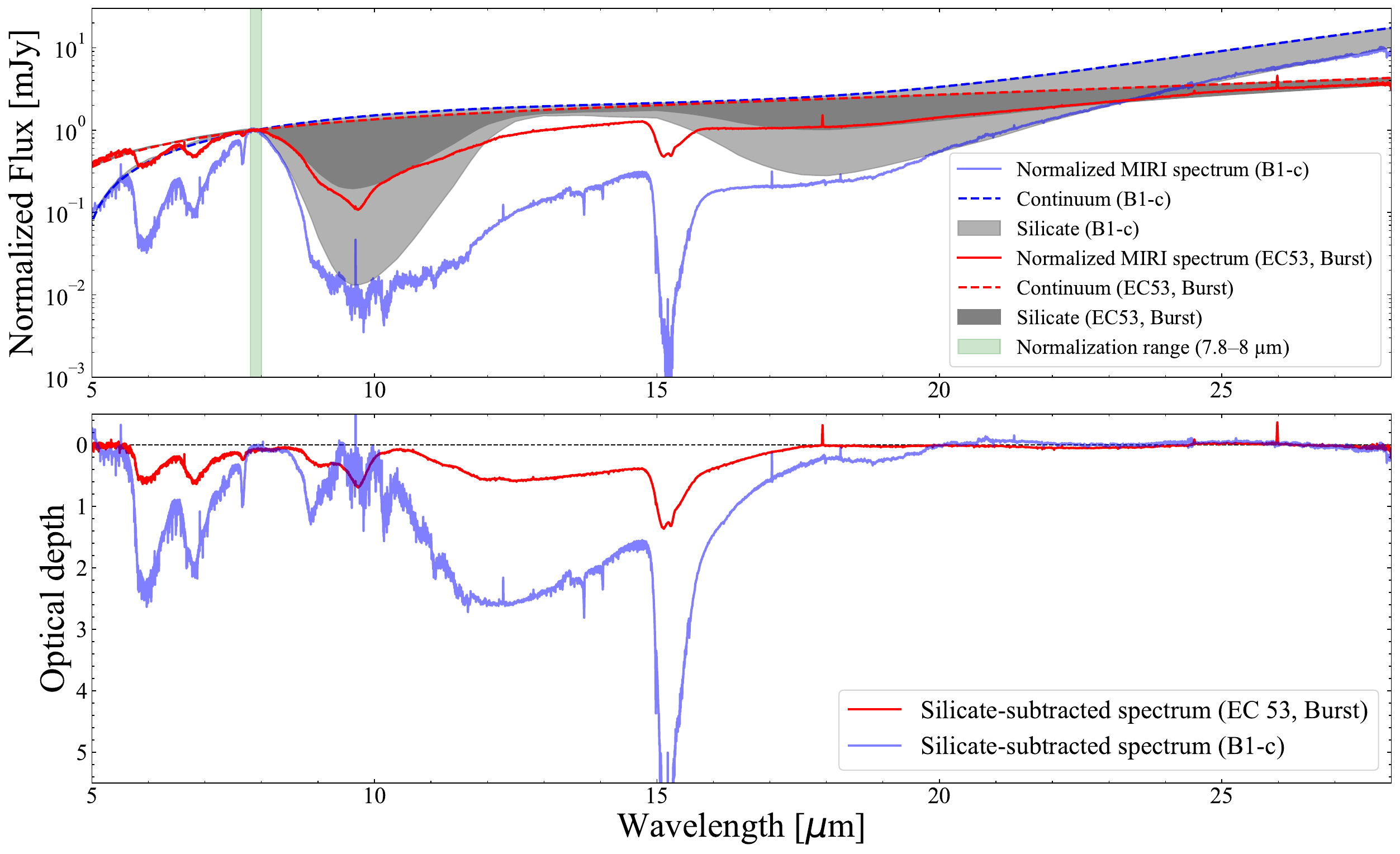}\
    \caption{
    Comparison of the continuum fitting and silicate subtraction applied to the JWST/MIRI spectra of EC~53 and B1-c (JWST GTO program 1290; PI: E.~F.~van Dishoeck). The top panel shows the normalized spectra (solid lines), best-fit continua (dashed lines), and fitted silicate absorption profiles (shaded regions) for both sources using an identical normalization range (7.8--\microns{8.0}) and fitting procedure.
    The bottom panel shows the resulting silicate-subtracted spectra in optical depth space.
    }
    \label{fig:compare_B1c}
\end{figure}

B1-c provides a valuable benchmark for this comparison because it exhibits prominent mid-infrared silicate absorption \citep{Boogert2008,chen2024} without previously reported high ice abundances, while having JWST observations of comparable spectral quality.
Applying the same continuum fitting and silicate subtraction procedure used for EC~53 to the JWST/MIRI MRS spectrum of B1-c obtained from the Guaranteed Time Observation (GTO) program (1290, PI: E. F. van Dishoeck) yields a markedly different outcome.
As shown in Figure~\ref{fig:compare_B1c}, the broad absorption centered near \microns{9.8} in B1-c is well reproduced by the silicate component, leaving only weak and localized residual ice absorption near \microns{9.7}, consistent with the CH$_3$OH stretching mode.
In contrast, EC~53 retains prominent ice absorption features across the same wavelength range even after subtraction of the silicate contribution.

This direct comparison demonstrates that the adopted analysis framework does not systematically enhance ice optical depths.
Instead, it effectively discriminates between silicate and ice absorption components.
The persistence of prominent ice residuals in EC~53, but not in B1-c, therefore indicates that the high ice abundances derived toward EC~53 are intrinsic to the source rather than artifacts of continuum fitting or silicate subtraction.

It is also instructive to compare our B1-c results with previous JWST analyses that adopted different continuum treatments.
\citet{chen2024} analyzed the \microns{9.8} absorption band in the MIRI MRS spectrum of B1-c using a local baseline fitting approach to isolate the \microns{9.7} absorption feature attributed to CH$_3$OH ice.
Despite these methodological differences, the isolated \microns{9.7} absorption inferred by \citet{chen2024} is in close agreement with the silicate-subtracted spectrum derived in our spectral analysis.

\begin{figure}[hp!]
    \centering
    \includegraphics[scale=0.25]{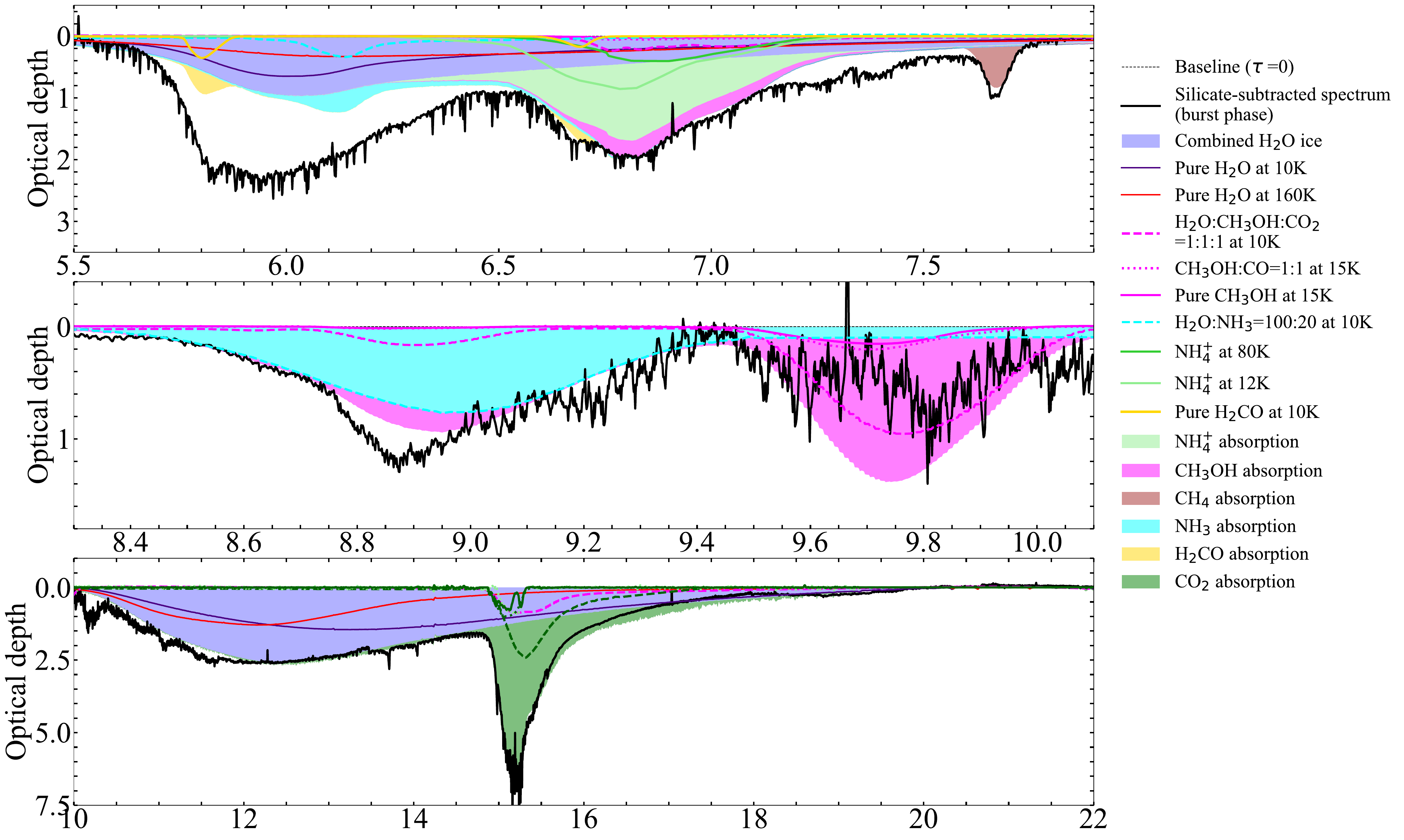}\
    \caption{
    Global ice composition of the major species applied to the silicate-subtracted JWST/MIRI MRS spectrum of B1-c.
    The shaded region shows the combined ice absorption constructed from the sum of laboratory profiles (solid and dashed lines) corresponding to individual ice species.
    The combined profile reproduces the major ice absorption bands across the 5--\microns{20} range, including contributions from H$_2$O (blue), CO$_2$ (green), CH$_3$OH (magenta), NH$_3$ (cyan), CH$_4$ (brown), NH$_4^+$ (light green), and H$_2$CO (gold).
    }
    \label{fig:B1c_global_fit}
\end{figure}

To further assess the robustness of our ice analysis and to place the EC~53 results in a quantitative context, we applied our global ice fitting procedure to the silicate-subtracted spectrum of B1-c, focusing on the major ice absorption bands. Figure~\ref{fig:B1c_global_fit} shows the resulting global ice fit, in which the major ice absorption bands are well reproduced using the same set of laboratory ice profiles adopted for EC~53. From this fit, we derived column densities and relative ice abundances for the major ice species toward B1-c. The resulting abundances of CH$_3$OH and CH$_4$ are in close agreement with those reported by \citet{chen2024}, demonstrating that our continuum determination, silicate subtraction, and global ice fitting framework robustly recovers ice abundances consistent with previous JWST studies, despite differences in methodology. 

At the same time, the ice abundances derived for B1-c remain significantly lower than those measured toward EC~53.
This contrast reinforces the conclusion that the exceptionally high ice abundances observed toward EC~53 are intrinsic to the source rather than an artifact of the analysis methodology.

\begin{deluxetable*}{l cccccccc c @{\hspace{0.35cm}} ccccc}
\tablecaption{Ice abundances with respect to H$_2$O \label{tbl:abundance}}
\tablewidth{0pt}
\setlength{\tabcolsep}{0.03in}

\tablehead{
  \colhead{} &
  \multicolumn{8}{c}{Robust detections} &
  \multicolumn{5}{c}{Tentative candidates} \\
  \cline{2-9}\cline{11-15}
  \colhead{Source} &
  \colhead{CO} &
  \colhead{CO$_2$} &
  \colhead{NH$_3$} &
  \colhead{CH$_4$} &
  \colhead{CH$_3$OH} &
  \colhead{H$_2$CO} &
  \colhead{NH$_4^+$} &
  \colhead{OCN$^-$} &
  \colhead{} &
  \colhead{CH$_3$CHO} &
  \colhead{CH$_3$CH$_2$OH} &
  \colhead{CH$_3$COOH} &
  \colhead{HCOOH} &
  \colhead{NH$_2$CHO }
}

\startdata
This work (EC~53)
  & 26.59 & 49.56 & 34.96 & 6.64 & 25.50 & 2.33 & 16.10 & 1.23 &
  & $<$0.97  & $<$5.03  & $<$1.16  & $<$2.81  & $<$2.69 \\
NGC 1333 IRAS 2A\tablenotemark{a}
  & {\it --}   & {\it --}   & {\it --}   & 1.6   & 5.0, 7.6 & 4.1  & {\it --}   & 1.2 &
  & {\it --} & 1.2 & 0.3 & 1.0 & $<$0.4 \\
Ced 110 IRS4A\tablenotemark{b}
  & 2.0   & 24.5  & 6.4   & 1.7   & 2.1      & $<$7.5 & $<$14.3 & 2.0 &
  & {\it --} & {\it --} & {\it --} & 2.0 & {\it --} \\
Ced 110 IRS4B\tablenotemark{b}
  & 10.4  & 11.2  & 8.9   & {\it --} & 6.1    & {\it --}  & {\it --}  & {\it --} &
  & {\it --} & {\it --} & {\it --} & {\it --} & {\it --} \\
B1-c\tablenotemark{c}
  & {\it --}   & {\it --}   & {\it --}   & 3.7   & 11.9     & 1.8  & {\it --}  & 1.6 &
  & 0.74  & 0.85  & $<$0.08 & 0.44 & {\it --} \\
LYSOs (Spitzer)\tablenotemark{d}
  & 21    & 28    & 6     & 4.5   & 6        & 6    & 11   & 0.6 &
  & {\it --} & {\it --} & {\it --} & 0.5$-$4 & $<$3.7\tablenotemark{e}
\enddata

\tablecomments{
All abundances are in percentiles (\%). Columns are grouped into robustly analyzed species (CO, CO$_2$, NH$_3$, CH$_4$, CH$_3$OH, H$_2$CO, NH$_4^+$, and OCN$^-$) and tentatively considered species (CH$_3$CHO, CH$_3$CH$_2$OH, CH$_3$COOH, HCOOH, and NH$_2$CHO ). We refer to the ice abundances of those four low-mass protostars conducted from the JWST programs (IceAge; \citet{McClure2023} and JOYS; \citet{vanDishoeck2025}).
}
\tablenotetext{a}{The ice abundances are taken from \citet{rocha2024}}
\tablenotetext{b}{The ice abundances are taken from \citet{Rocha2025}}
\tablenotetext{c}{The ice abundances are taken from \citet{chen2024}}
\tablenotetext{d}{The ice abundances are taken from \citet{Boogert2015}}
\tablenotetext{e}{\citet{Slavicinska2023}}
\end{deluxetable*}

\begin{deluxetable}{lcccc}
\tablecaption{Ice abundances toward B1-c compared with previous JWST analysis\label{tbl:B1c_abundance}}
\tablehead{
\colhead{Species} &
\colhead{This work} &
\colhead{Chen et al. (2024)} &
\colhead{EC~53} &
\colhead{$\lambda_{peak}$ ($\mu$m)}
}
\startdata
CH$_3$OH & 10.3 & 11.9 & 25.50 & \microns{9.74} \\
CH$_4$   & 4.1  & 3.7  & 6.64  & \microns{7.67} \\
CO$_2$   & 35.1 & ---  & 49.56 & \microns{15.25} \\
NH$_3$   & 15.3 & ---  & 34.96 & \microns{9.0} \\
NH$_4^+$  & 10.5 & ---  & 16.10 & \microns{6.85} \\
H$_2$CO  & 2.1  & 1.8  & 2.33  & \microns{8.02} \\
\enddata
\tablecomments{
Abundances are given as percentages relative to H$_2$O.
Values from this work are derived from the global ice fit applied to the silicate-subtracted JWST/MIRI MRS spectrum of B1-c.
For reference, the total H$_2$O ice column density derived in this work
($N(H_2O) = 2.28 \times 10^{19}~\mathrm{cm^{-2}}$)
is consistent within uncertainties with the H$_2$O column density reported by \citet{chen2024}, despite differences in continuum treatment and fitting methodology.
Only major ice species quantitatively analyzed by \citet{chen2024} are directly compared.
}
\end{deluxetable}

\section{Summary\label{sec:summary}}

In this study, we present a comprehensive and self-consistent infrared spectroscopic analysis of the Class \RomanNumeralCaps{1} protostar EC~53 (V371 Ser) using JWST NIRSpec and MIRI observations obtained in both quiescent and burst phases. 
Our workflow (i) determines a robust continuum for NIRSpec with a hybrid polynomial+GPR scheme, (ii) performs a simultaneous continuum and silicate absorption fit for MIRI, and (iii) performs a global ice decomposition analysis across the near- to mid-infrared spectral range with laboratory profiles using Bayesian MCMC.

Although the ice absorption spectra show rich and complex absorption features corresponding to major ice species including H$_2$O, CO$_2$, CO, CH$_3$OH, NH$_3$, CH$_4$, as well as several minor COMs, the shapes and depths of those ice bands are indistinguishable within uncertainties between epochs. We interpret that moderate and short-duration accretion bursts in EC~53 do not appreciably restructure the line-of-sight ice reservoir.

The envelope hosts an ice-rich and compositionally complex mantle: relative to H$_2$O, the major ices show elevated fractions (notably CO$_2$, CH$_3$OH, NH$_3$) compared to typical embedded protostars, pointing to efficient cold-phase surface chemistry prior to protostellar heating. The large fraction of pure NH$_3$ compared to its H$_2$O-rich counterpart further supports an extended cold history.

A prominent double-peaked CO$_2$ bending mode at \microns{15.2}, best reproduced by CDE-corrected pure CO$_2$, indicates segregation/distillation from CO-rich mixtures at $\sim$20$-$30~K. The H$_2$O ice composition requires a mix of 10~K amorphous and 160~K crystalline components, signaling partial annealing. The broad NH$_4^+$ absorption near \microns{6.8} (with a temperature-dependent redshift) and the distinct OCN$^-$ feature at \microns{4.62} indicate acid–base and radical chemistry activated by thermal and energetic processing.
These features point to episodic heating consistent with accretion variability. 

Therefore, the ice composition of EC~53 might reflect a two-phase evolutionary path.
A long, cold prestellar phase that enabled the accumulation of chemically mature ices via hydrogenation pathways.
Subsequent episodic heating history, likely tied to accretion bursts, introduced thermal and energetic processing, shaping both absorption characteristics and complexity.
The coexistence of cold-phase products (e.g., pure NH$_3$) with thermally evolved components (NH$_4^+$, segregated CO$_2$, and crystalline H$_2$O) favors a layered structure in which deeper, shielded strata preserve pristine materials, while outer layers record intermittent heating.
This interpretation makes EC~53 a compelling case for understanding how variable accretion modulates the chemical makeup of icy mantles in low-mass star formation, bridging the gap between inherited interstellar ices and processed circumstellar materials.

\begin{acknowledgments}
This work was supported by the National Research Foundation of Korea (NRF) grant funded by the Korea government (MSIT) (grant numbers RS-2024-00416859 and RS-2026-25490557). D.J. is supported by NRC Canada and by an NSERC Discovery Grant. G.J.H. is supported by grant IS23020 from the Beijing Natural Science Foundation. G.B. was supported by Basic Science Research Program through the National Research Foundation of Korea (NRF) funded by the Ministry of Education (RS-2023-00247790). Y.A. acknowledges support by JSPS KAKENHI grant no. 24K00674. Y.-L.Y. acknowledges support from Grant-in-Aid from the Ministry of Education, Culture, Sports, Science, and Technology of Japan (20H05844 and 25H00676). J.D.G. acknowledges support from the associated 3477 NASA observer grant. Part of this research was carried out at the Jet Propulsion Laboratory, California Institute of Technology, under a contract with the National Aeronautics and Space Administration (80NM0018D0004). 

This work is based on observations made with the NASA/ESA/CSA James Webb Space Telescope. The data were obtained from the Mikulski Archive for Space Telescopes at the Space Telescope Science Institute, which is operated by the Association of Universities for Research in Astronomy, Inc., under NASA contract NAS 5-03127 for JWST. These observations are associated with JWST program \#3477. Some of the data presented in this article were obtained from the Mikulski Archive for Space Telescopes (MAST) at the Space Telescope Science Institute. The specific observations analyzed can be accessed via \dataset[doi:10.17909/h56s-xr27]{https://doi.org/10.17909/h56s-xr27}.
We acknowledge the use of ChatGPT for checking English grammar and improving the clarity of expressions.
\end{acknowledgments}
\facilities{JWST}

\begin{appendix} \label{sec:appendix}
\renewcommand\thefigure{\Alph{section}\arabic{figure}}
\setcounter{figure}{0}

\section{Optimization process of the laboratory ice absorbance profiles\label{sec:appendix_A}}
We collect laboratory infrared spectra of pure and mixed ice species from the Leiden Ice Database for Astrochemistry (LIDA; \citealt{Rocha2022}) and the NASA Goddard Cosmic Ice Library (\citealt{Hudson1999}). Both pure ice components and various H$_2$O-rich ice mixtures were considered, given their relevance in astrophysical environments. 
Laboratory spectra are typically provided at multiple temperatures; however, since the EC~53 protostellar envelope is expected to exhibit ice temperatures predominantly between 10 and 20~K, we selected spectra measured at 10~K or, if unavailable, at the next lower temperature.

Before applying these laboratory spectra to our global ice decomposition analysis, careful baseline correction and spectral optimization are essential. Some of the selected laboratory profiles underwent a baseline-correction procedure to match the optical-depth baseline of the silicate-subtracted spectrum of EC~53.
The baseline of pure H$_2$O ice absorption at 10~K was corrected by fitting the laboratory spectrum around the wavelength ranges of 2.4-2.7, 3.35-4.0, 4.8-5.3, 8.8-10, and 20-\microns{22} with a sixth-order polynomial function. For the crystalline pure H$_2$O ice at 160~K, we used its original spectrum, which more accurately represents the baseline of the optical depth spectrum than the 10~K spectrum. We also performed a baseline correction for the H$_2$O-rich ice mixture containing the CH$_3$OH, CH$_4$, and HCOOH profiles, avoiding the corresponding absorption bands of the ice species, based on the fitting points of the pure H$_2$O ice profile.
NH$_3$ ice was only used for the baseline correction among the pure ice species. We corrected the baseline of the pure NH$_3$ ice profile using the fitting ranges of 2.0-2.9, 3.2-5.9, 6.5-8.5, and beyond \microns{10}.

\begin{figure}[hp!]
    \centering
    \includegraphics[scale=0.3]{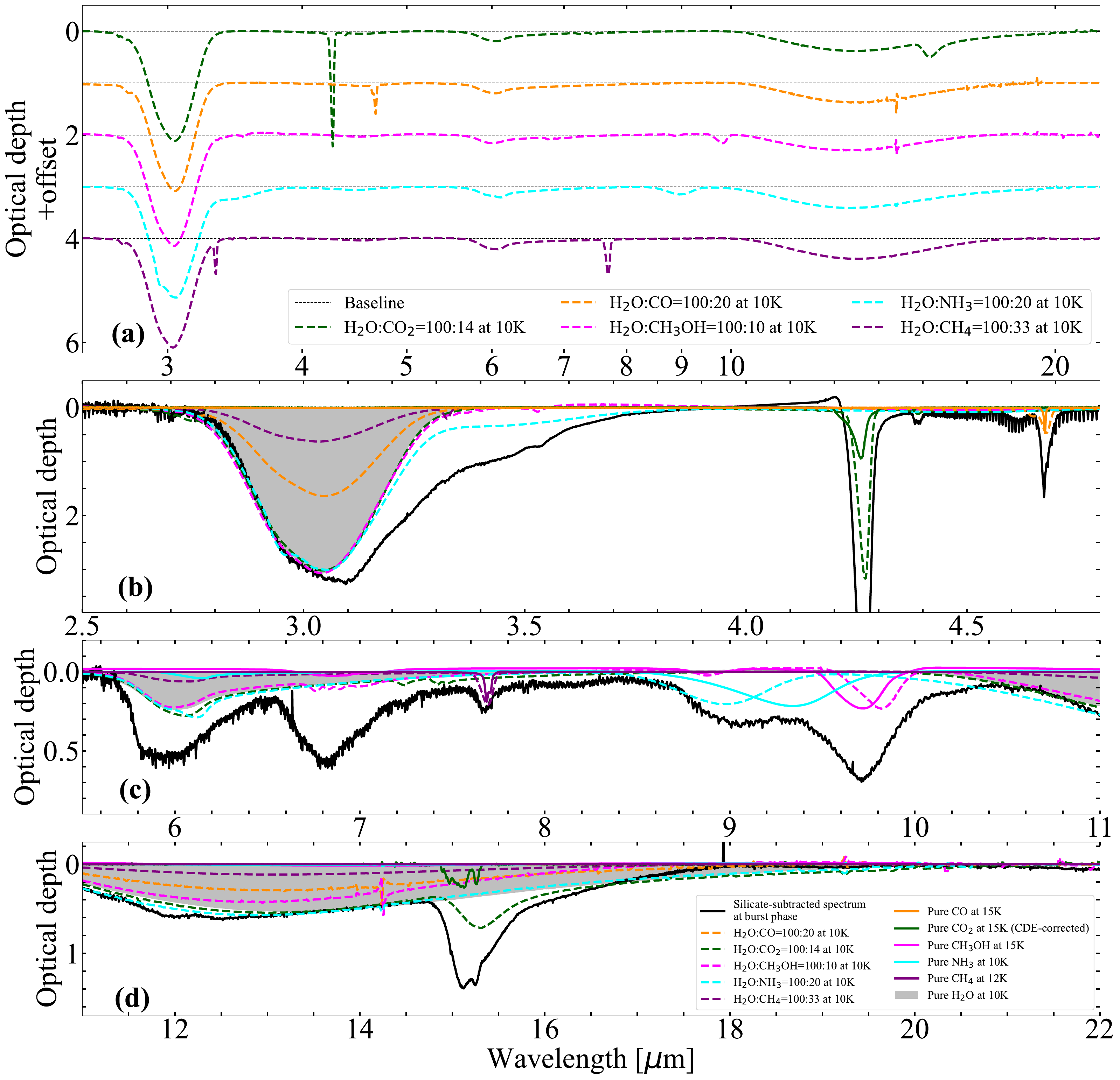}\
    \caption{(a) Baseline-corrected laboratory spectra of ice mixtures with H$_2$O at 10 K, including CO$_2$, CO, CH$_3$OH, NH$_3$, and CH$_4$. (b–d) Application of these optimized profiles to the silicate-subtracted spectrum of EC 53 during its burst phase, highlighting representative ice absorption bands. Comparisons between pure ice species and their corresponding H$_2$O-rich mixtures clearly illustrate notable differences in absorption profiles, underscoring the importance of water-rich mixtures for accurately reproducing observed spectral features.}
    \label{fig_appx:Ice_Lab_optim}
\end{figure}

Figure~\ref{fig_appx:Ice_Lab_optim} illustrates examples of baseline-corrected laboratory spectra for several representative ice mixtures with H$_2$O at 10 K, including CO$_2$, CO, CH$_3$OH, NH$_3$, and CH$_4$. Panel (a) shows the corrected and offset laboratory spectra, and panels (b–d) demonstrate their applications to the observed JWST spectrum during the burst phase, clearly highlighting differences between pure and H$_2$O-rich mixture profiles. This figure emphasizes the need to use appropriate mixture profiles to correctly identify and quantify ice components in complex astrophysical environments.

\begin{figure}[hp!]
    \centering
    \includegraphics[scale=0.3]{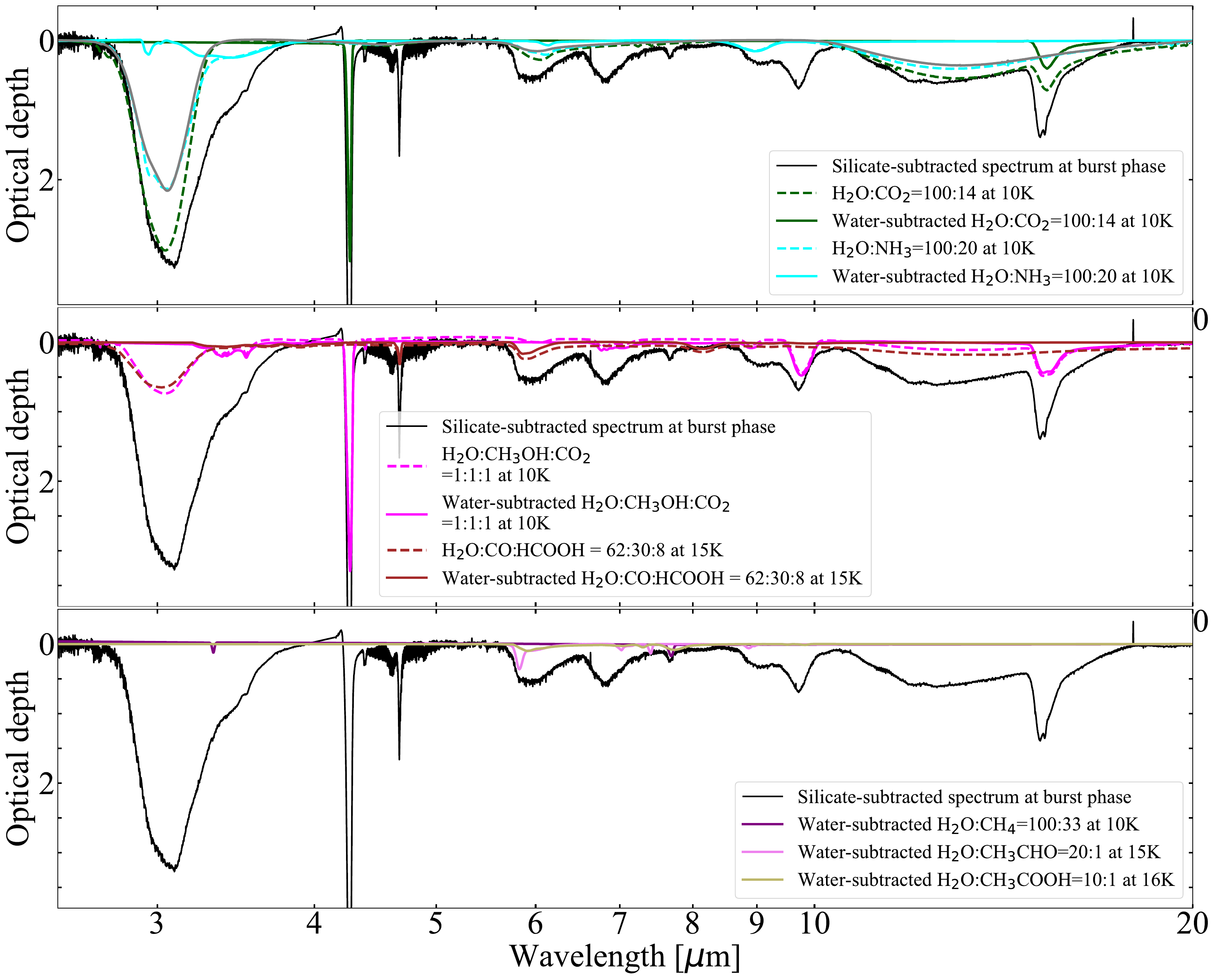}\
    \caption{Laboratory ice profiles of the H$_2$O-rich mixture optimized for the global ice analysis applied to the silicate-subtracted optical depth spectrum of EC 53 at the burst phase. This is an Illustration of the water-subtraction procedure applied to H$_2$O-rich ice mixtures used in the global ice-decomposition analysis. Each panel compares the original laboratory spectra of H$_2$O-rich mixtures (dashed lines) with their corresponding water-subtracted spectra (solid lines), highlighting isolated absorption contributions from embedded ice species after removing the dominant water-ice component (gray solid line in top panel). Ice mixtures shown include: (top) H$_2$O:CO$_2$ (100:14, 10 K) and H$_2$O:NH$_3$ (100:20, 10 K); (middle) H$_2$O:CH$_3$OH:CO$_2$ (1:1:1, 10 K) and H$_2$O:CO:HCOOH (62:30:8, 15 K); (bottom) H$_2$O:CH$_4$ (100:33, 10 K), H$_2$O:CH$_3$CHO (20:1, 15 K), and H$_2$O:CH$_3$COOH (10:1, 16 K). This subtraction approach effectively isolates the modified absorption profiles of minor ice species, eliminating interference from the broad and intense absorption bands of water ice. All spectral profiles were baseline-corrected and interpolated to match the JWST spectral resolution, as detailed in Appendix~\ref{sec:appendix_A}.}
    \label{fig_appx:Ice_Lab_H2O_subt}
\end{figure}

Furthermore, water-rich ice mixtures inherently contain substantial contributions from broad, dominant H$_2$O absorption features, which can obscure or distort the characteristic absorption bands of embedded ice species. To accurately isolate absorption features from minor components in water-rich mixtures, we systematically removed the intrinsic H$_2$O contribution from these mixture profiles. The detailed process is illustrated in Figure~\ref{fig_appx:Ice_Lab_H2O_subt}. The original laboratory spectra of representative ice mixtures are shown as dashed lines, and their water-subtracted versions as solid lines. Specifically, we demonstrate this subtraction for mixtures such as H$_2$O:CO$_2$, H$_2$O:NH$_3$, H$_2$O:CH$_3$OH:CO$_2$, H$_2$O:CO:HCOOH, H$_2$O:CH$_4$, H$_2$O:CH$_3$CHO, and H$_2$O:CH$_3$COOH. The resulting water-subtracted profiles clearly reveal the spectral shapes unique to each minor species, enabling more precise spectral fitting and identification of these components within the observed EC 53 spectrum.

\setcounter{figure}{0}

\section{Detailed process of the global ice fit \label{sec:appendix_B}}
In this section, we present the detailed global ice analysis procedure used to sequentially fit the absorption features in the EC~53 spectrum, decomposing them into individual ice species based on their pure and mixed laboratory profiles. To optimally match the observed absorption features with laboratory ice profiles, we employed Bayesian inference using Markov Chain Monte Carlo (MCMC) sampling implemented with the \texttt{emcee} Python package \citep{Foreman-Mackey2013}. This approach allowed us to derive stable, physically consistent scale coefficients for each ice component.

We began with a silicate-subtracted optical depth spectrum ($\tau_{\rm obs}(\lambda)$), ensuring residual absorption features primarily originated from ice species. We modeled the absorption region using a linear combination of laboratory ice profiles listed in Table~\ref{tbl:lab_ref}:
\begin{equation}
 \tau_{\rm model}(\lambda;\mathbf{a})
  = \sum_{i=1}^{n} a_i \; \tau_{{\rm lab},i}(\lambda), \label{eq:1}\\\end{equation}
where $a_i\ge0$ represents the scale coefficient for each laboratory ice profile $i$, and $\tau_{{\rm lab},i}(\lambda)$ denotes its optical depth.
Assuming independent Gaussian noise with variance $\sigma^{2}$, estimated via adjacent-point differencing of $\tau_{\rm obs}(\lambda)$, we defined the log-likelihood function as:
\begin{equation}
\ln p\!\bigl(\tau_{\mathrm{obs},j}\mid \mathbf{a}\;\tau_{\mathrm{lab},j}\bigr)
  = -\frac{1}{2} \sum_{j}
    \frac{\bigl(\tau_{{\rm obs},j}-\sum_i a_i\tau_{{\rm lab},i,j}\bigr)^2}
         {\sigma^2},    \label{eq:2}
\\\end{equation}
where $j$ denotes indices covering the fitted wavelength range.
Posterior distributions of these scaling coefficients were determined using ensemble sampling with 50 walkers over 3,000 steps. Median values provided the best-fit scale coefficients, with uncertainties derived from the 16th and 84th percentiles, approximating one standard deviation (1$\sigma$) confidence intervals.

Column density distributions were obtained by integrating scaled laboratory optical-depth profiles over wavelength intervals defined by the respective absorption features. 
For each component $i$ and each MCMC draw $\mathbf{a}^{(k)}$, column densities were calculated as:
\begin{equation}
    N_i^{(k)}
  = \frac{1}{A_i}
    \int_{\lambda_{\min,i}}^{\lambda_{\max,i}}
      a_i^{(k)}\;\tau_{{\rm lab},i}(\lambda)
    \;d\tilde\nu,
  \quad
  \tilde\nu \equiv \frac{1}{\lambda\,10^{-4}}\quad(\mathrm{cm}^{-1}),\label{eq:3}
\\\end{equation}
where \(A_i\) is the band strength (cm molecule\(^{-1}\); Table~\ref{tbl:lab_ref}) and the integral is performed in the wavenumber space.

Subsequent subsections detail the global fitting results for each step, focusing on specific wavelength regions and illustrating confidence correlations via corner plots of the derived column densities.

\subsection{Step 1. H$_2$O ice composition \label{subsec:appendix_B1}}
\begin{figure}[h!]
    \centering
    \includegraphics[scale=0.23]{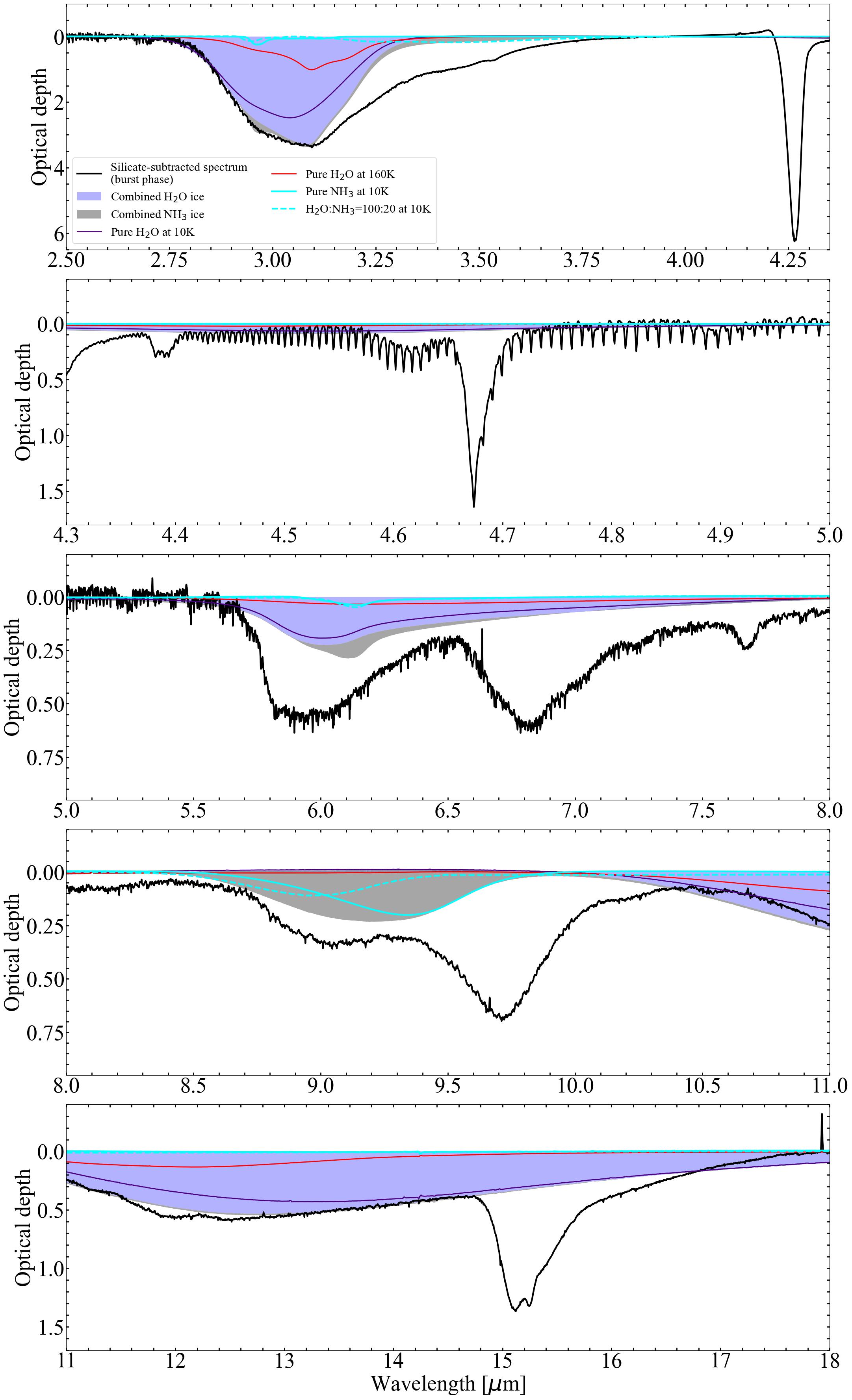}\
    \caption{First step of the global ice-fitting process for EC 53. The combined H$_2$O~ice absorption (blue-shaded region) is composed of laboratory profiles of pure H$_2$O ice measured at temperatures of 10~K (purple) and 160~K (red). Also shown are contributions from NH$_3$-based ices, including pure NH$_3$ at 10~K (solid cyan) and an H$_2$O-rich NH$_3$ ice mixture at 10~K (dashed cyan). This initial step isolates the dominant H$_2$O\ absorption features, notably at the \microns{3}, \microns{4.5}, \microns{6}, and \microns{13} vibrational bands, providing baseline constraints essential for accurate decomposition of subsequent ice components.}
    \label{fig_appx:Ice_fit_step1}
\end{figure}

As described in section ~\ref{sec:results}, the initial step of the global fitting procedure focused on analyzing the ice composition of H$_2$O, which dominates the absorption features in the EC~53 spectrum. The MCMC fit was performed primarily around the prominent \microns{3} band, where absorption by H$_2$O~ice is strongest. In addition, minor contributions from NH$_3$-based ice profiles (pure and H$_2$O-rich mixtures) were considered within the narrower spectral range of 2.7-\microns{3.1}. These profiles were further cross-checked against their absorption around the \microns{9} region, where their signatures are more distinctly evident. The best-fit results confirmed that the combined H$_2$O ice profiles accurately reproduced the libration mode absorption centered near \microns{13}, demonstrating consistency across multiple vibrational modes (Figure~\ref{fig_appx:Ice_fit_step1}).
This step provided baseline constraints for subsequent processes.

\begin{figure}[hp]
    \centering
    \includegraphics[scale=0.5]{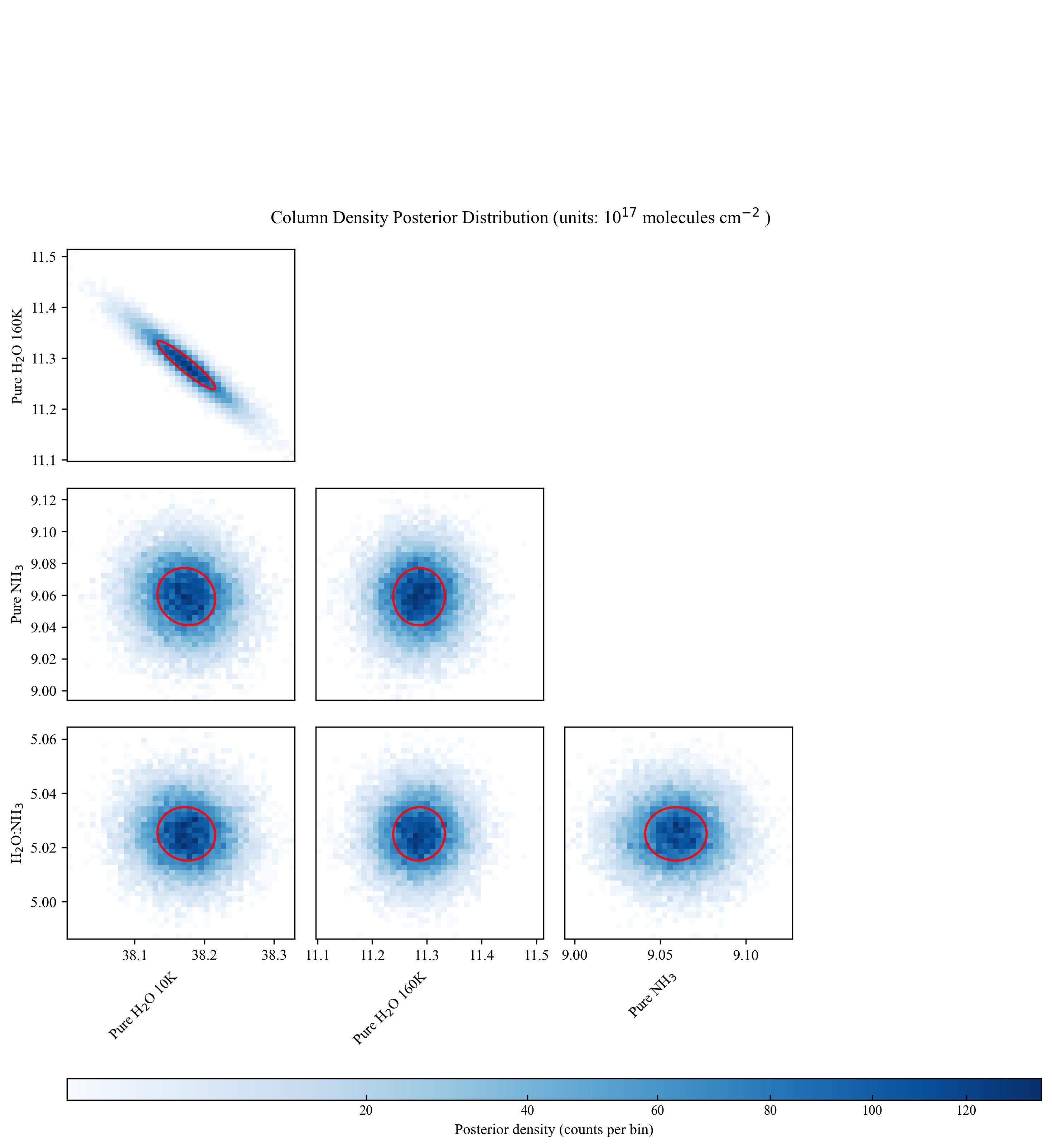}\
    \caption{Corner plot displaying posterior distributions and correlations for the column densities of the H$_2$O and NH$_3$-based ice components derived from the global ice fit around the \microns{3} absorption band. The distributions illustrate the 1$\sigma$ (68\%) confidence boundary (red circles) of the column density estimates.}
    \label{fig_appx:Ice_fit_step1_colden}
\end{figure}

\subsection{Step 2. CO$_2$ ice composition \label{subsec:appendix_B2}}
\begin{figure}[hp]
    \centering
    \includegraphics[scale=0.4]{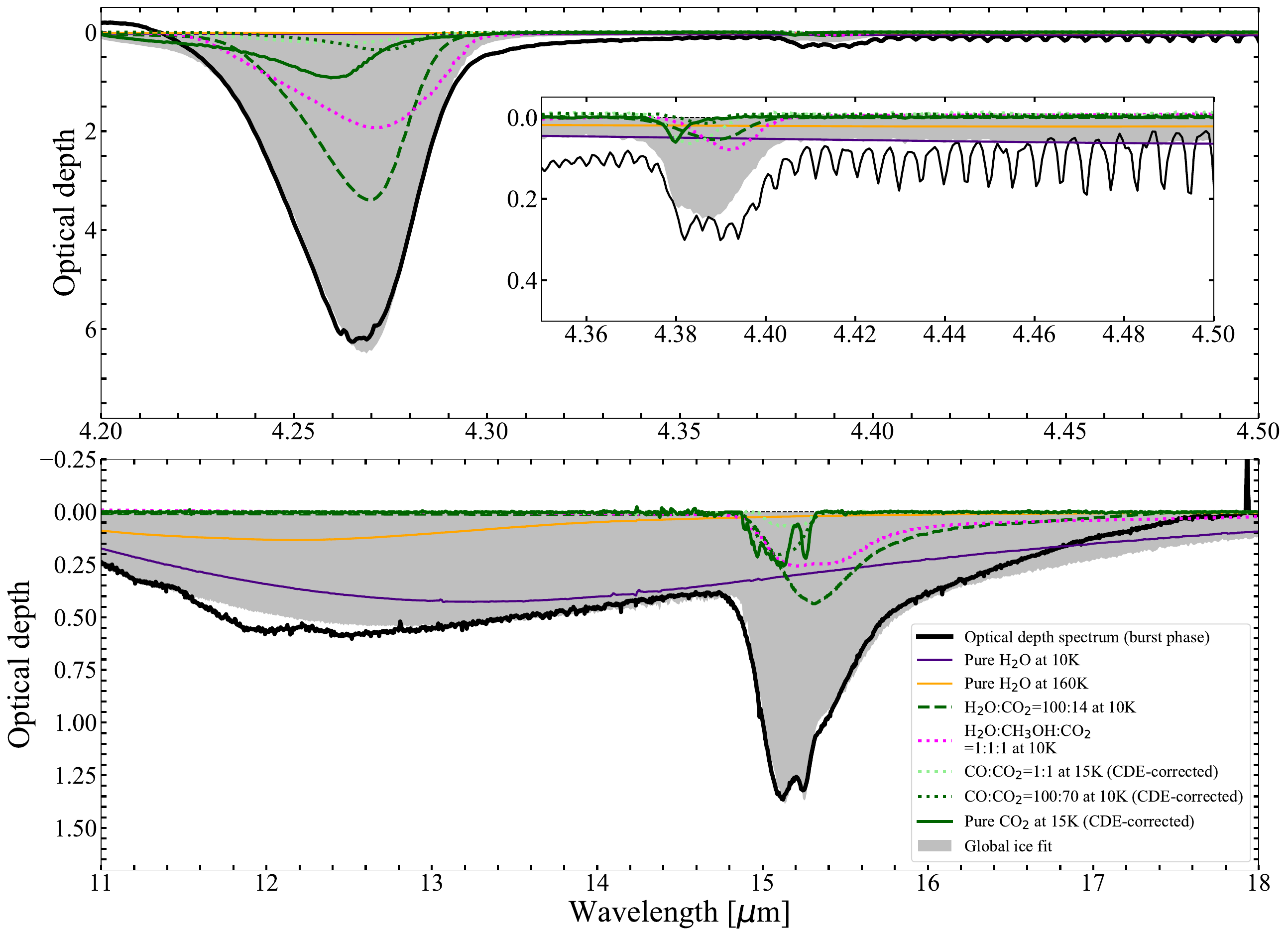}\
    \caption{Global ice decomposition of the silicate-subtracted JWST optical depth spectrum of EC~53 during the burst phase, focusing on spectral contributions from CO$_2$ ice. 
    \textit{Upper panel (4.2-\microns{4.5})}: Fundamental stretching mode of CO$_2$ ice is shown, highlighting contributions from pure CO$_2$ at 15~K (CDE-corrected; solid dark green line), mixtures of H$_2$O:CO$_2$ (100:14 at 10~K; dashed dark green line), H$_2$O:CH$_3$OH:CO$_2$ (1:1:1 at 10~K; magenta line), and CDE-corrected CO-mixed CO$_2$ profiles (CO:CO2 ratios of 1:1 at 15~K and 100:70 at 10~K; dotted light and dark green lines). The inset emphasizes the subtle yet diagnostic isotopic absorption feature of $^{13}$CO$_2$ around \microns{4.38}.
    \textit{Lower panel (11-\microns{18})}: Corresponding bending mode of CO$_2$ ice near \microns{15.2} is illustrated. The fitted CO$_2$ ice composition, which integrates all laboratory ice profiles, is depicted by the gray-shaded region. Dominant contributions from pure amorphous (10~K; purple) and crystalline (160~K; orange) H$_2$O ice profiles are also indicated.}
    \label{fig_appx:Global_fit_CO2}
\end{figure}

\begin{figure}[ht]
    \centering
    \includegraphics[scale=0.23]{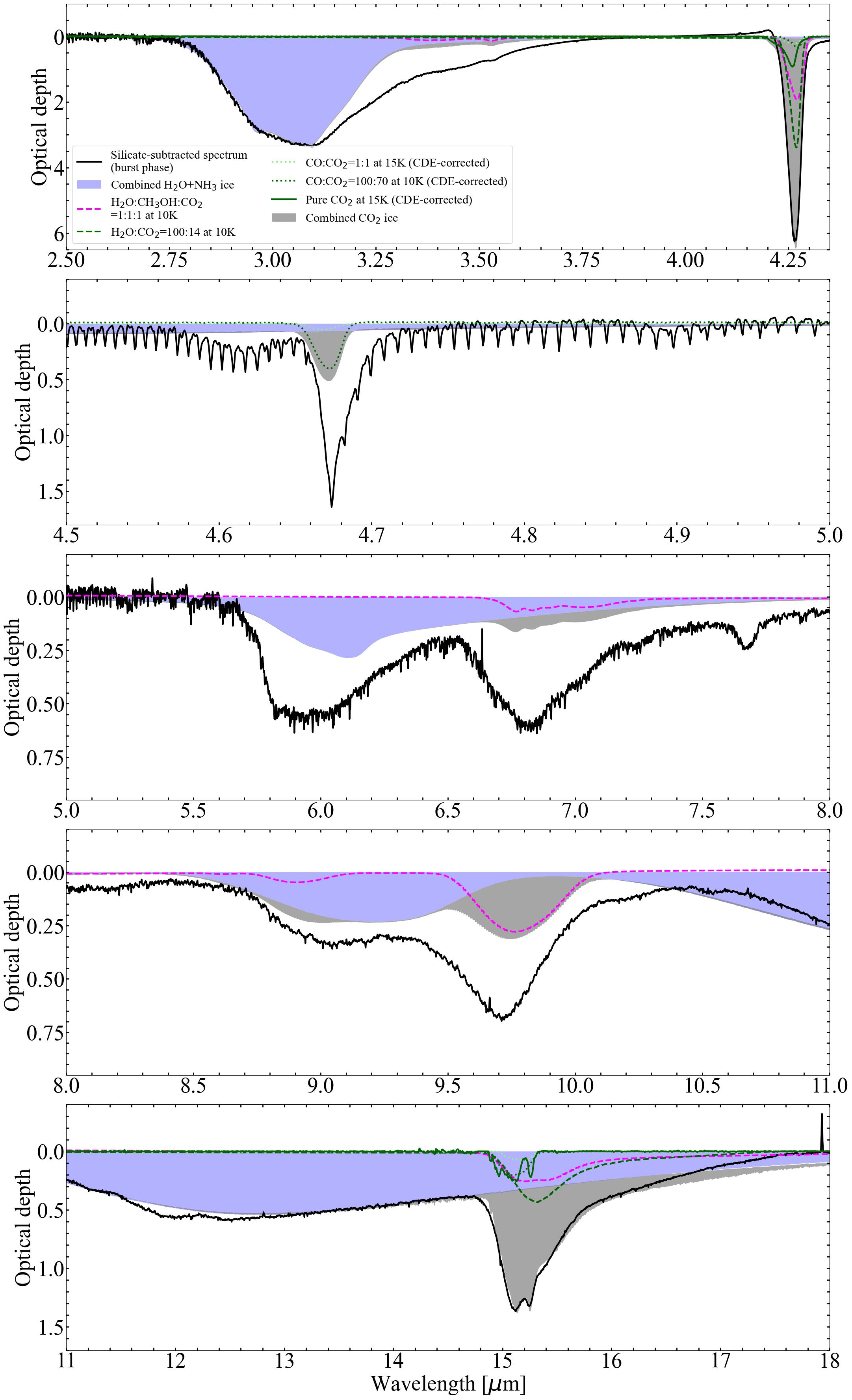}\
    \caption{Second step of the global ice-fitting process for EC~53, highlighting contributions from CO$_2$-based ice components. The combined absorption from H$_2$O and NH$_3$-based ice profiles derived in the first step is shown by the blue-shaded region.
    The laboratory ice profiles used for fitting the CO$_2$ absorption features were initially scaled using results from the preliminary fitting depicted in Figure~\ref{fig_appx:Global_fit_CO2}.
    These initial values were refined via MCMC fitting, yielding the corrected global composition of CO$_2$ ice shown here.}
    \label{fig_appx:Ice_fit_step2}
\end{figure}

In this step, guided by the H$_2$O ice absorption previously fitted in Step 1, we simultaneously analyzed the CO$_2$-based ice absorption features centered around the \microns{4.27} and \microns{15} bands, as well as the isotopic $^{13}$CO$_2$ absorption at \microns{4.38}. To accurately reproduce the distinct double-peaked structure at \microns{15.2}, we employed a CDE-corrected pure CO$_2$ ice profile measured at 15~K. Additionally, to represent both the central and blue-side absorption structures of the \microns{15} band and the subtle isotopic feature at \microns{4.38}, we included CDE-corrected laboratory profiles of CO-mixed CO$_2$ ices with CO:CO$_2$ ratios of 1:1 (at 15~K) and 100:70 (at 10~K). Furthermore, a CH$_3$OH-mixed CO$_2$ ice profile (H$_2$O:CH$_3$OH:CO$_2$ = 1:1:1 at 10~K) was incorporated to better fit the absorption features of the red side of these bands. Given the challenge of clearly distinguishing contributions from the H$_2$O-rich CO$_2$ ice mixture (H$_2$O:CO$_2$ = 100:14 at 10~K) independently, the initial fitting procedure was performed manually by simultaneously adjusting the contribution from CH$_3$OH ice, especially focusing on its absorption near \microns{9.7}. The initial scaling of these laboratory profiles was guided by preliminary fitting results (Figure~\ref{fig_appx:Global_fit_CO2}), which were refined through the MCMC fit. The final corrected global composition of the CO$_2$ ice derived from this approach is presented in Figure~\ref{fig_appx:Ice_fit_step2}.

\begin{figure}[ht!]
    \centering
    \includegraphics[scale=0.5]{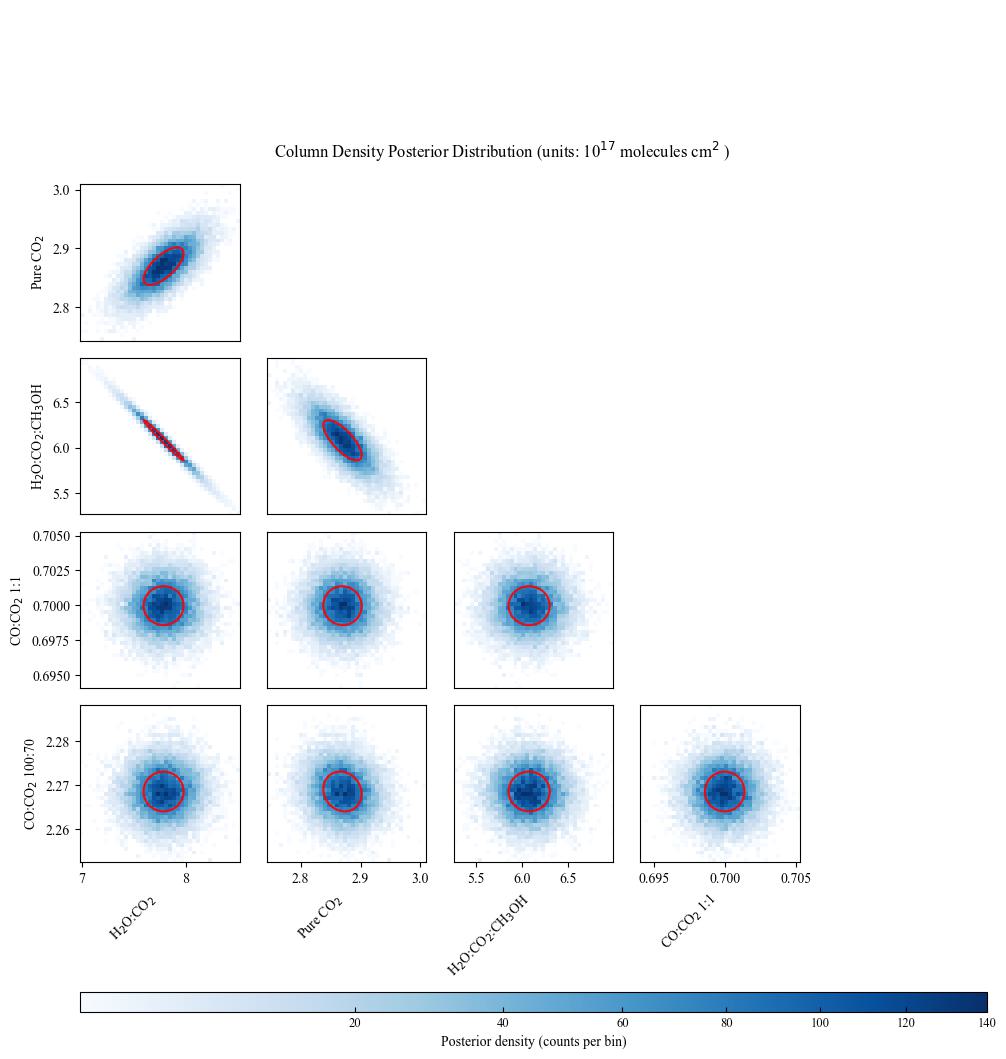}\
    \caption{Corner plot displaying posterior distributions of column densities for CO$_2$-based ice components derived from the global ice fit around the \microns{15} absorption band. The red circles indicate the 1$\sigma$ (68\%) confidence intervals of the column density estimates, highlighting correlations and statistical uncertainties among the fitted ice components.}
    \label{fig_appx:Ice_fit_step2_colden}
\end{figure}
\clearpage
\subsection{Step 3. CH$_3$OH ice composition and refinement ice fit to the \microns{9} absorption band \label{subsec:appendix_B3}}
Based on the intensity of the H$_2$O:CH$_3$OH:CO$_2$ = 1:1:1 ice profile determined in Step 2, we refined the global ice composition by analyzing the absorption features around the prominent \microns{9.7} band, primarily attributed to the CH$_3$OH-bearing ices. However, since this spectral region partially overlaps with the absorption of NH$_3$-based ice components, both the pure and H$_2$O-rich NH$_3$ ice profiles derived from Step 1 were simultaneously included in the MCMC fitting procedure. 

\begin{figure}[h!]
    \centering
    \includegraphics[scale=0.23]{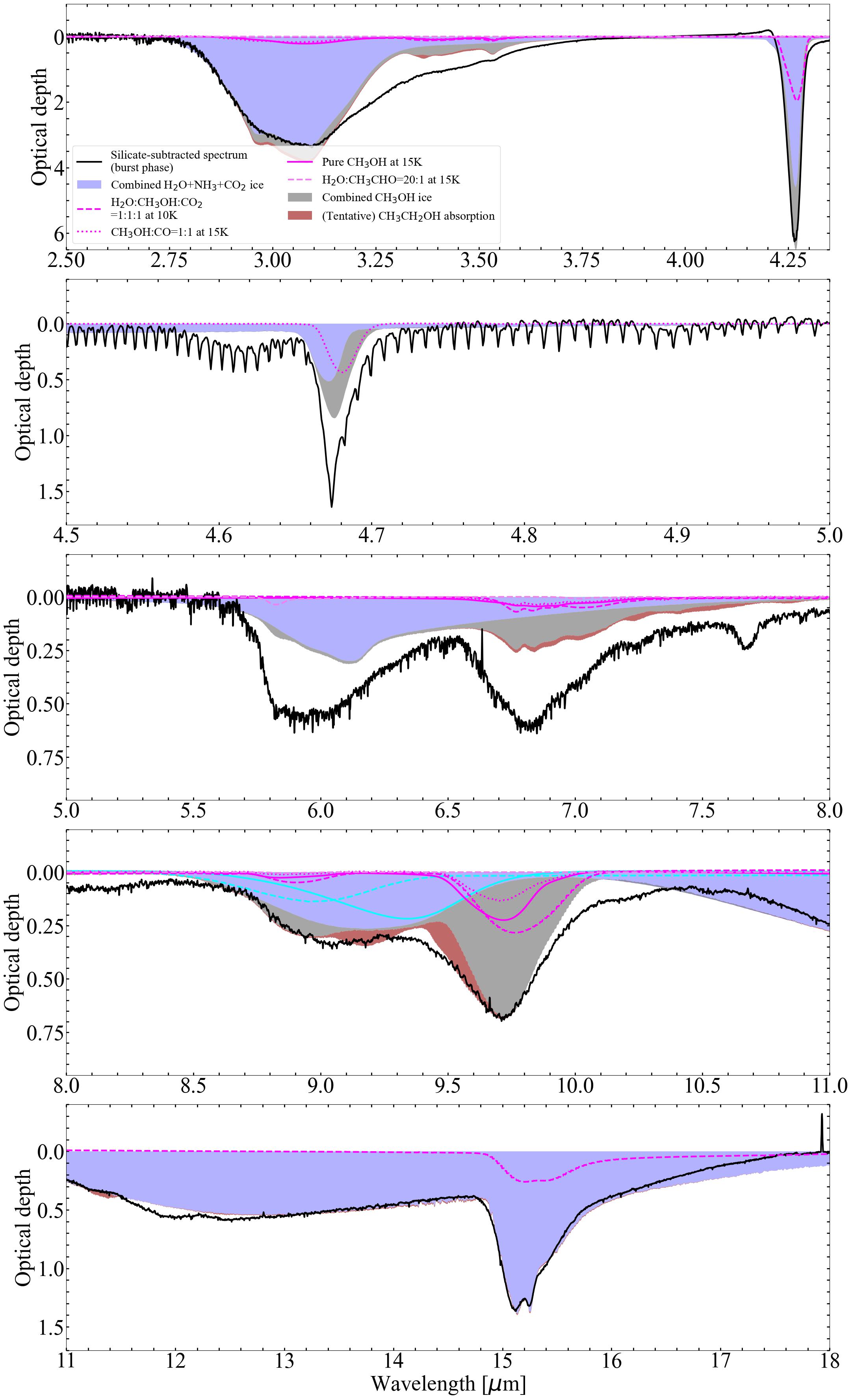}\
    \caption{Third step of the global ice-fitting process for EC~53, focusing on the contributions from CH$_3$OH-bearing ice components. The combined ice absorption derived from Step 2 is shown as the blue-shaded region. Overlaid are laboratory profiles used to model CH$_3$OH-based ice absorption: pure CH$_3$OH at 15~K (solid magenta), a CO-mixed CH$_3$OH profile (CO:CH$_3$OH = 1:1 at 15~K, dotted magenta), and a ternary mixture profile of H$_2$O:CH$_3$OH:CO$_2$ (1:1:1 at 10~K, dashed magenta). Additionally, tentative contributions from other minor ice species such as CH$_3$CH$_2$OH (brown-shaded region) and an H$_2$O-rich CH$_3$CHO mixture (20:1 at 15~K, dashed violet) are included. This step refines the fit of the CH$_3$OH ice absorption bands across the 3.5-\microns{10} range, particularly optimizing the fit around the \microns{9.7} band, and provides a corrected scaling for the previously fitted CH$_3$OH-CO$_2$ mixture profile at both \microns{15} and \microns{9.7} bands.}
    \label{fig_appx:Ice_fit_step3}
\end{figure}

\begin{figure}[ht!]
    \centering
    \includegraphics[scale=0.5]{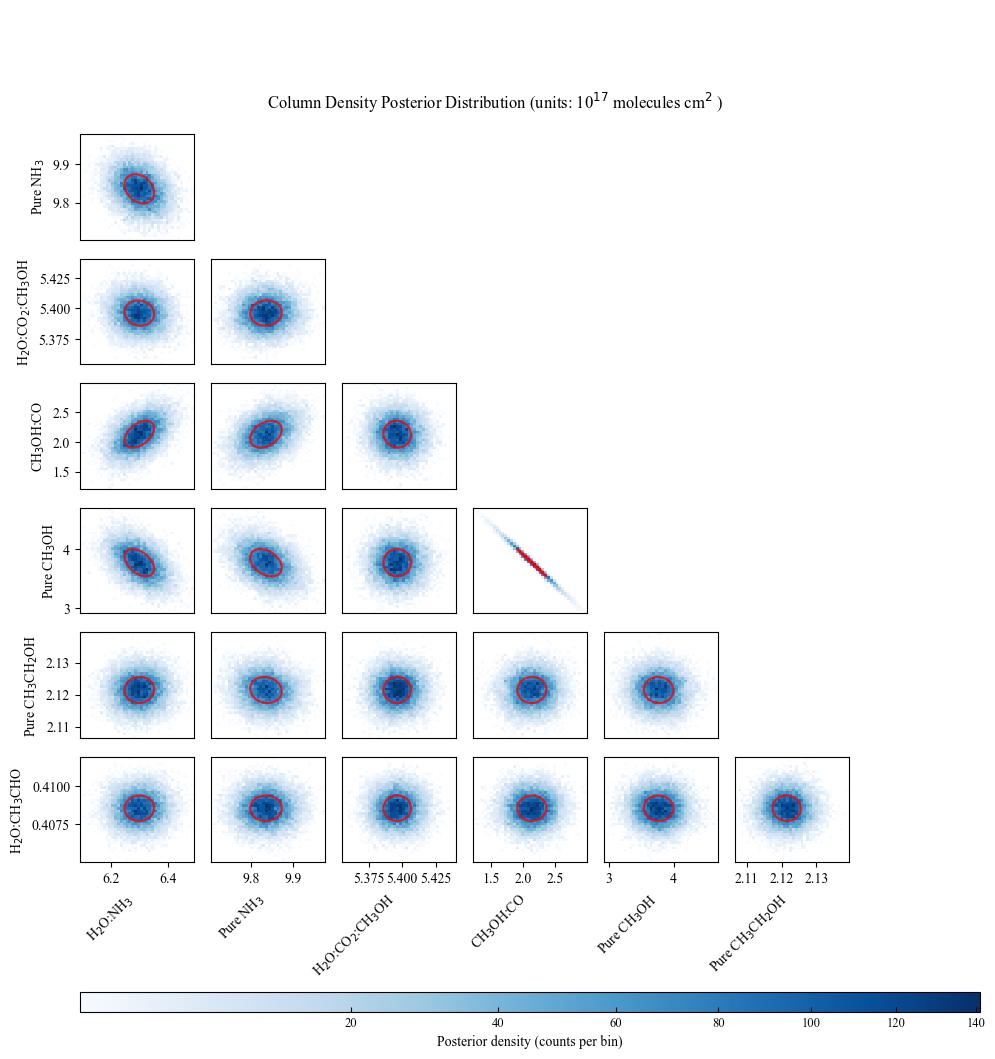}\
    \caption{Corner plot illustrating posterior distributions of column densities for the CH$_3$OH-based ice components derived from the global ice fit around the \microns{9.7} absorption band. The red cirlces represent the 1$\sigma$ (68\%) confidence intervals, clearly visualizing correlations and statistical uncertainties among the various ice components, including pure and mixed forms of CH$_3$OH, as well as other minor contributing species such as NH$_3$, CH$_3$CH$_2$OH, and CH$_3$CHO.}
    \label{fig_appx:Ice_fit_step3_colden}
\end{figure}

We observed that an overestimation of the pure NH$_3$ ice component could lead to excessive absorption in its stretching mode near \microns{2.9}, inconsistent with the observed spectrum. To mitigate this risk, we tentatively included a pure CH$_3$CH$_2$OH ice profile, known to exhibit characteristic absorption peaks around \microns{9.17} and \microns{9.5}. Although clear evidence of this species was not strongly present in the spectrum, its potential contribution was considered to prevent artificial enhancement of pure NH$_3$ ice absorption. Similarly, to accurately model subtle absorption around \microns{8.9}, we included an H$_2$O-rich CH$_3$CHO mixture (H$_2$O:CH$_3$CHO = 20:1 at 15~K), even though its contribution to the final fit was minor. The resulting global ice composition after these adjustments is shown in Figure~\ref{fig_appx:Ice_fit_step3}, with the statistical confidence intervals of column densities presented in Figure~\ref{fig_appx:Ice_fit_step3_colden}.

Despite these considerations, distinguishing clearly between pure and CO-mixed CH$_3$OH ice absorption profiles remained challenging when based solely on the \microns{9.7} band analysis. 
Therefore, to better clarify the relative proportions of these two ice components, we further analyzed the CO-mixed CH$_3$OH ice profile in conjunction with the clearly distinguishable CO ice absorption feature at \microns{4.67}, in the subsequent fitting step.

\clearpage

\subsection{Step 4. CO and OCN$^-$ ice composition \label{subsec:appendix_B4}}
In this step, we analyzed the prominent absorption feature of CO ice centered around \microns{4.67}. The fitting components considered for this spectral region included not only pure CO ice but also various mixtures previously constrained in earlier steps, such as CO$_2$-mixed, CH$_3$OH-mixed, and H$_2$O-rich CO (H$_2$O:CO:HCOOH = 62:30:8 at 15~K) mixtures. Additionally, due to the proximity and partial overlap of absorption features, the laboratory profile of OCN$^-$ ice at 12~K, centered at \microns{4.62}, was simultaneously incorporated into the MCMC fitting procedure.

\begin{figure}[hp!]
    \centering
    \includegraphics[scale=0.23]{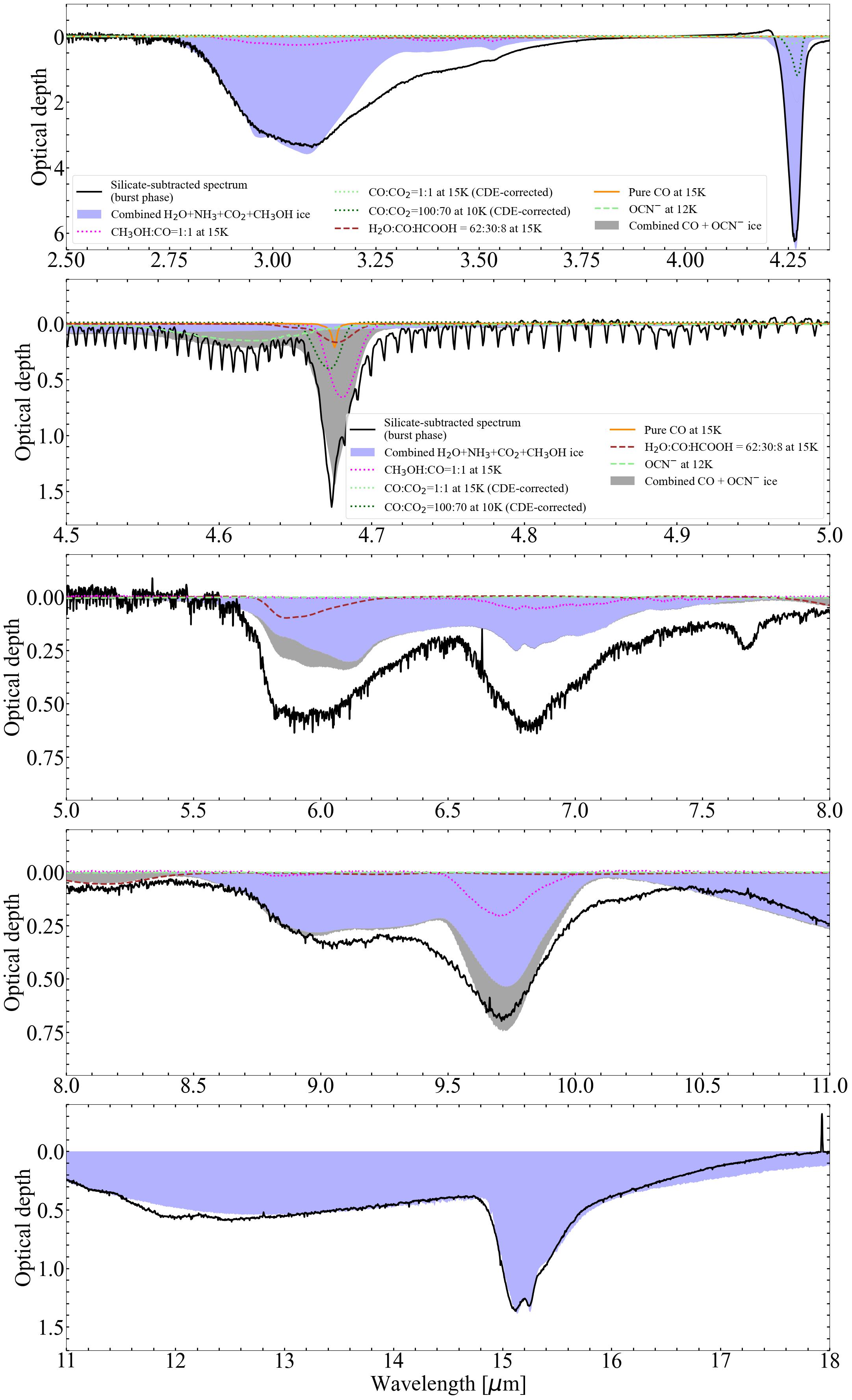}\
    \caption{Fourth step of the global ice-fitting process for EC~53, highlighting contributions from CO-bearing ice components. The combined absorption profile from H$_2$O, NH$_3$, CO$_2$, and CH$_3$OH-based ices derived from the previous step is represented by the blue-shaded region. Overlaid are laboratory profiles of various CO-based ice components, including pure CO ice at 15~K (solid orange), a CO-mixed CH$_3$OH ice profile (CO:CH$_3$OH = 1:1 at 15~K, dotted magenta), CO-mixed CO$_2$ profiles (CO:CO$_2$ ratios of 1:1 at 15~K and 100:70 at 10~K, dotted dark and light green), and a ternary ice mixture of H$_2$O:CO:HCOOH (62:30:8 at 15~K, dashed brown). Additionally, the absorption due to OCN$^-$ at \microns{4.62}, an indicator of energetic processing, is fitted using the corresponding laboratory ice profile measured at 12~K (dashed light green). This step isolates the spectral contributions of CO-bearing ices around the \microns{4.67} band, allowing refined determination of both the CH$_3$OH-mixed CO ice profile and the associated absorptions at the \microns{4.67} and \microns{9.7} bands.}
    \label{fig_appx:Ice_fit_step4}
\end{figure}

\begin{figure}[ht!]
    \centering
    \includegraphics[scale=0.5]{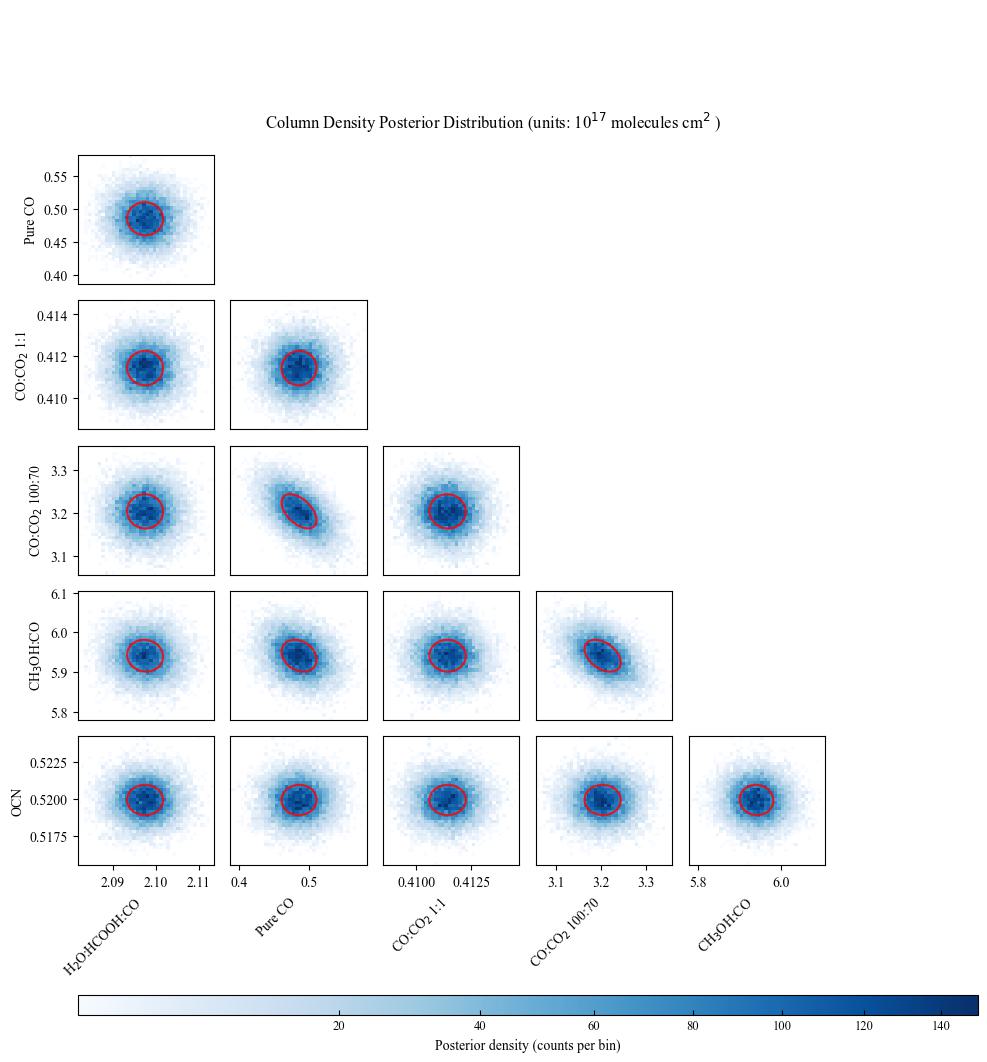}\
    \caption{Corner plot displaying posterior distributions of column densities for the CO-based ice components derived from the global ice fit around the \microns{4.67} absorption band, including the closely neighboring OCN$^-$ ice absorption at \microns{4.62}. The red circles represent the 1$\sigma$ (68\%) confidence intervals, illustrating statistical uncertainties and correlations among the column densities of the pure CO, CO-mixed ice components, and OCN$^-$ ices.}
    \label{fig_appx:Ice_fit_step4_colden}
\end{figure}

The initial scaling for the CO$_2$-mixed CO ice profiles, established in Step 2, served as the starting values for the MCMC fitting. The initial scaling of the H$_2$O-rich CO ice mixture was specifically adjusted to account for absorption by HCOOH ice observed around \microns{8}. Figure~\ref{fig_appx:Ice_fit_step4} illustrates the resulting global fit and shows that the combined absorption intensities of these CO-based ice components provide an excellent match to the spectral feature observed in \microns{4.67}. Notably, this fitting result indicated that the CO-mixed CH$_3$OH ice component (CH$_3$OH:CO = 1:1 at 15 K) contributes significantly more to the overall absorption at \microns{9.7} than pure CH$_3$OH ice. Consequently, to refine the composition ratio of these CH$_3$OH-based ice components, we performed an additional detailed analysis of the \microns{9.7} absorption region in the subsequent step. The statistical confidence intervals of the column densities for each CO-bearing ice component, including the OCN$^-$ ice absorption, are presented in Figure~\ref{fig_appx:Ice_fit_step4_colden}.

\subsection{Step 5. Refinement of CH$_3$OH-based ice composition \label{subsec:appendix_B5}}
In this refinement step, we revisited the global fitting of the CH$_3$OH-based ice absorption at the \microns{9.7} band. Based on updated constraints from previous steps, particularly the corrected intensity of the CO-mixed CH$_3$OH (CO:CH$_3$OH=1:1 at 15~K) profile derived from Step 4, we used these refined values as initial conditions for the MCMC fitting procedure. 

\begin{figure}[hp!]
    \centering
    \includegraphics[scale=0.23]{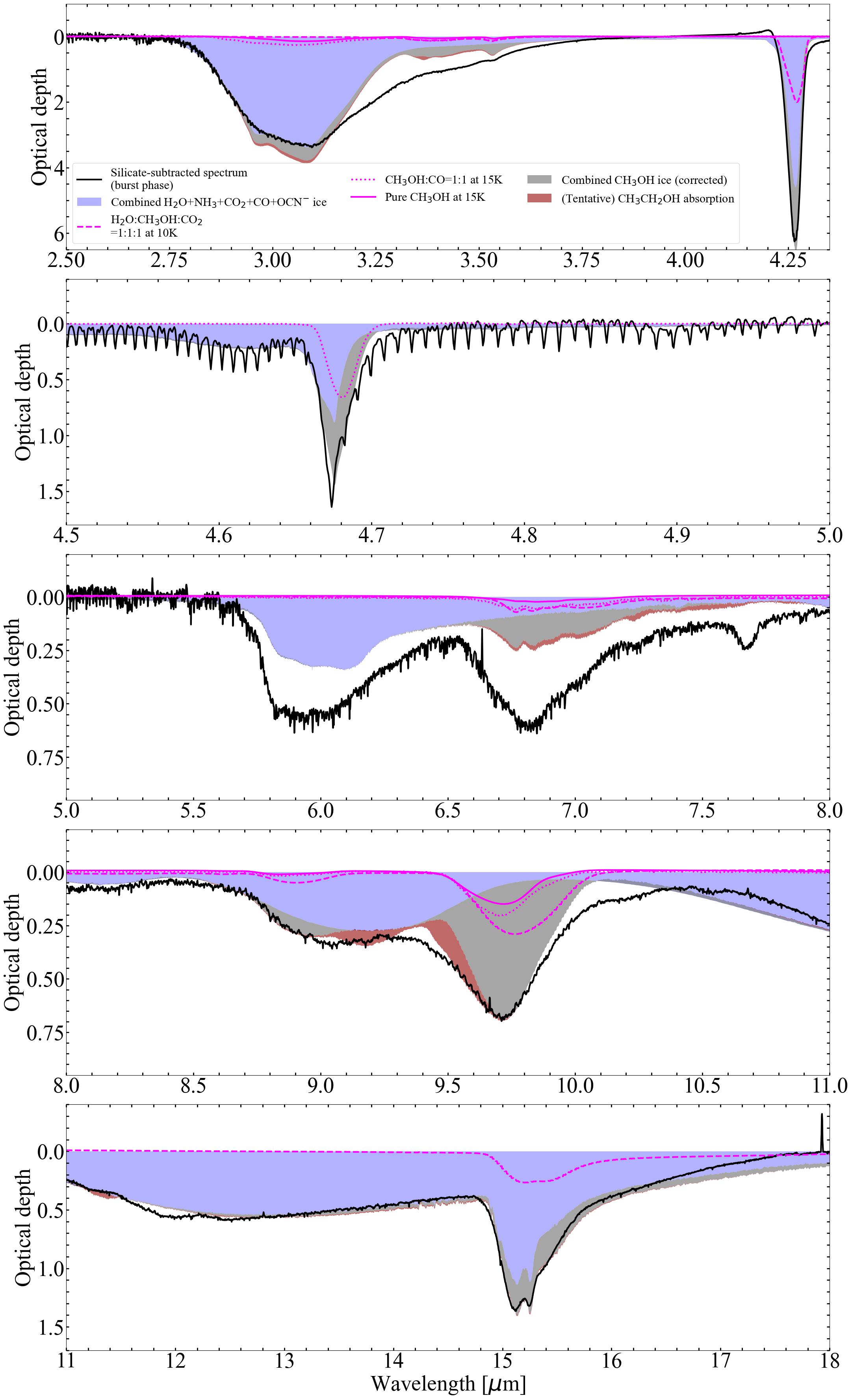}\
    \caption{Corrected global ice-fitting result of CH$_3$OH-bearing ice components (Step 5). This step reanalyzes and refines the previous fitting of the CH$_3$OH-based ice absorption around the prominent \microns{9.7} band by incorporating updated constraints from the Step 4 analysis, particularly concerning the CO-mixed CH$_3$OH component. The corrected combined CH$_3$OH ice absorption profile (gray-shaded region) demonstrates improved consistency across relevant spectral regions, notably the \microns{4.67} band, dominated by CO-mixed CH$_3$OH absorption, and the \microns{15.2} band, attributed primarily to CH$_3$OH-mixed CO$_2$ ice absorption.}
    \label{fig_appx:Ice_fit_step5}
\end{figure}

\begin{figure}[hp!]
    \centering
    \includegraphics[scale=0.5]{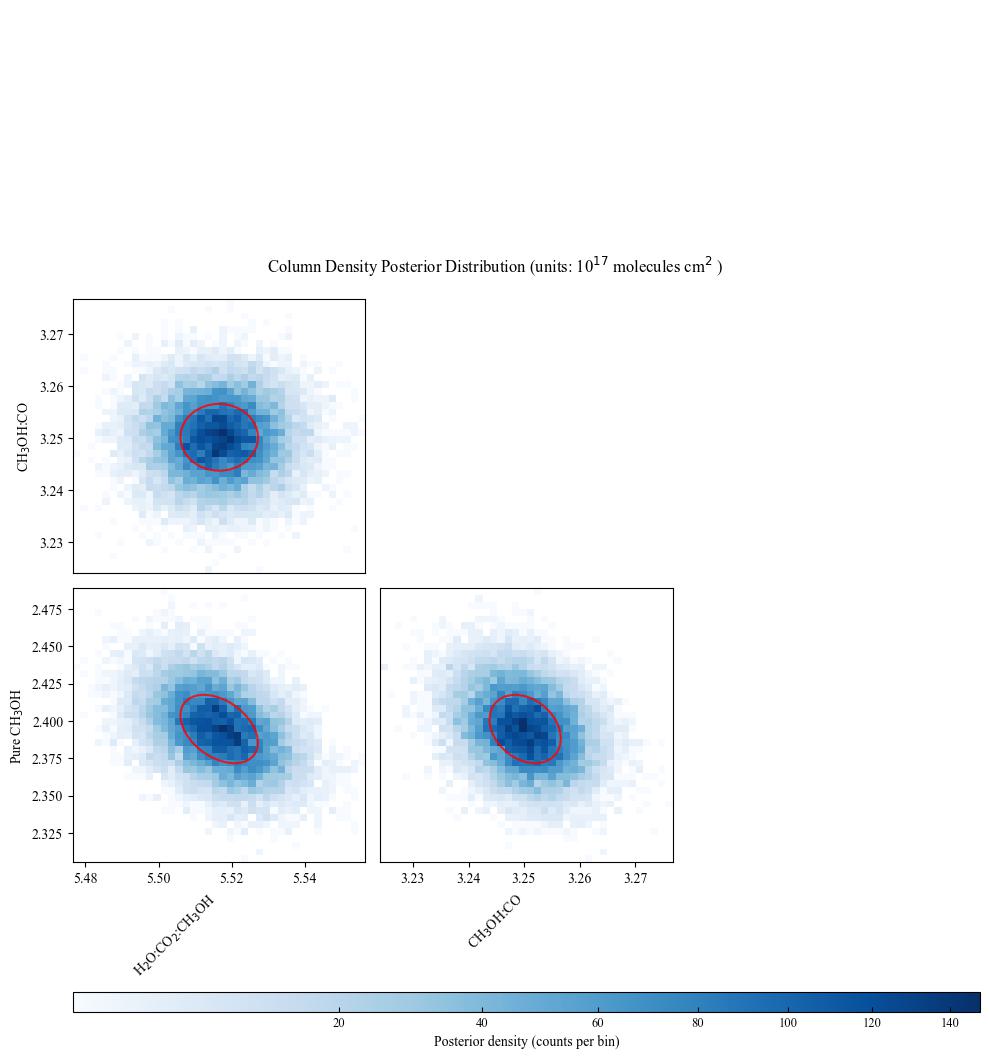}\
    \caption{Corner plot displaying corrected posterior distributions and confidence intervals for column densities of the CH$_3$OH-based ice components, reanalyzed and optimized around the \microns{9.7} band in Step 5. The red circles represent the 1$\sigma$ (68\%) confidence intervals, illustrating improved constraints on the relative abundances and uncertainties between pure CH$_3$OH, the CO-mixed CH$_3$OH, and the ternary H$_2$O:CH$_3$OH:CO$_2$ mixture, following the refined analysis of the CO ice feature at \microns{4.67} from the preceding step.}
    \label{fig_appx:Ice_fit_step5_colden}
\end{figure}

Figure~\ref{fig_appx:Ice_fit_step5} suggests that the corrected combination of these CH$_3$OH-bearing ice components achieves an improved fit across relevant spectral regions, especially around the \microns{4.67} band (associated with the CO-mixed CH$_3$OH component) and the \microns{15.2} band (attributed to the CH$_3$OH-mixed CO$_2$ component). As a result of this refined analysis, the relative contribution of pure CH$_3$OH ice absorption was substantially reduced, yielding a more physically consistent global ice composition. The final confidence intervals and posterior distributions of the column densities for each of these ice components are presented in Figure~\ref{fig_appx:Ice_fit_step5_colden}.

\subsection{Step 6. Ice composition within the 6-\microns{8} absorption region \label{subsec:appendix_B6}}
In this step, we analyzed ice components responsible for absorption features within the spectral region of 6 to \microns{8}, with particular emphasis on prominent absorption bands near \microns{6.8} and \microns{7.7}. For the \microns{6.8} band, laboratory profiles of NH$_4^+$ ice measured at 12~K and 80~K were used as primary fitting components. Additionally, the distinct absorption feature near \microns{6.7}, attributed to pure H$_2$CO ice (10~K), was included to further refine the spectral fitting within this range.

To represent the absorption centered at \microns{7.7}, we employed a laboratory profile of an H$_2$O-rich CH$_4$ ice mixture (H$_2$O:CH$_4$ = 100:33 at 10~K). Furthermore, to adequately model the absorption extending from approximately 7.7 to \microns{8}, we introduced a laboratory profile of an H$_2$O-rich CH$_3$COOH mixture (H$_2$O:CH$_3$COOH = 10:1 at 16~K). The spectral region between 7.0 and \microns{7.7} contains complex overlapping absorption features from various complex organic molecules (COMs), making it difficult to clearly isolate individual contributions. Consequently, we tentatively included additional laboratory profiles for pure NH$_2$CHO  ice, which could account for absorption features in both the 5.5-\microns{6.0} region and the complex 7.0-\microns{7.7} range. Another tentative profile, pure CH$_3$CH$_2$OH ice, was also considered to further improve the overall fit.

\begin{figure}[hp!]
    \centering
    \includegraphics[scale=0.23]{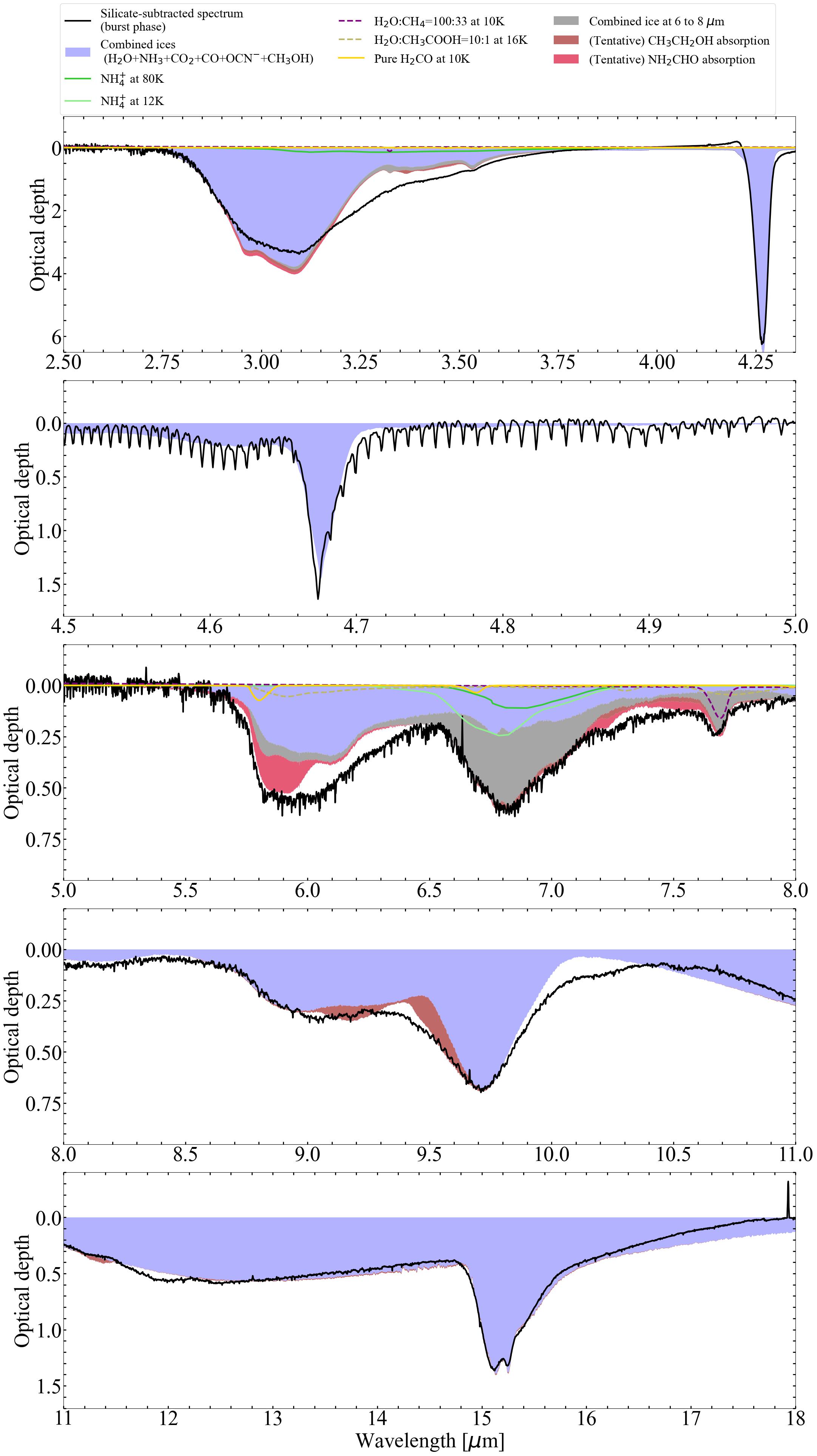}\
    \caption{Sixth step of the global ice-fitting process for EC~53, addressing ice absorption features within the 6–8 µm spectral range. The combined ice absorption profile (blue-shaded region) derived from Step 5 is shown for context. Overlaid are laboratory ice profiles used for fitting: NH$_4^+$ ice at 12~K (solid light green) and 80~K (solid green), pure H$_2$CO ice at 10~K (solid gold), and H$_2$O-rich mixtures of CH$_4$ (H$_2$O:CH$_4$ = 100:33 at 10~K, dashed purple) and CH$_3$COOH (H$_2$O:CH$_3$COOH = 10:1 at 16~K, dashed khaki). Additionally, tentative contributions from potential minor ice components, such as pure NH$_2$CHO  (crimson-shaded region) and CH$_3$CH$_2$OH (brown-shaded region), are included to improve the spectral fit to subtle absorption features.}
    \label{fig_appx:Ice_fit_step6}
\end{figure}

\begin{figure}[hp!]
    \centering
    \includegraphics[scale=0.5]{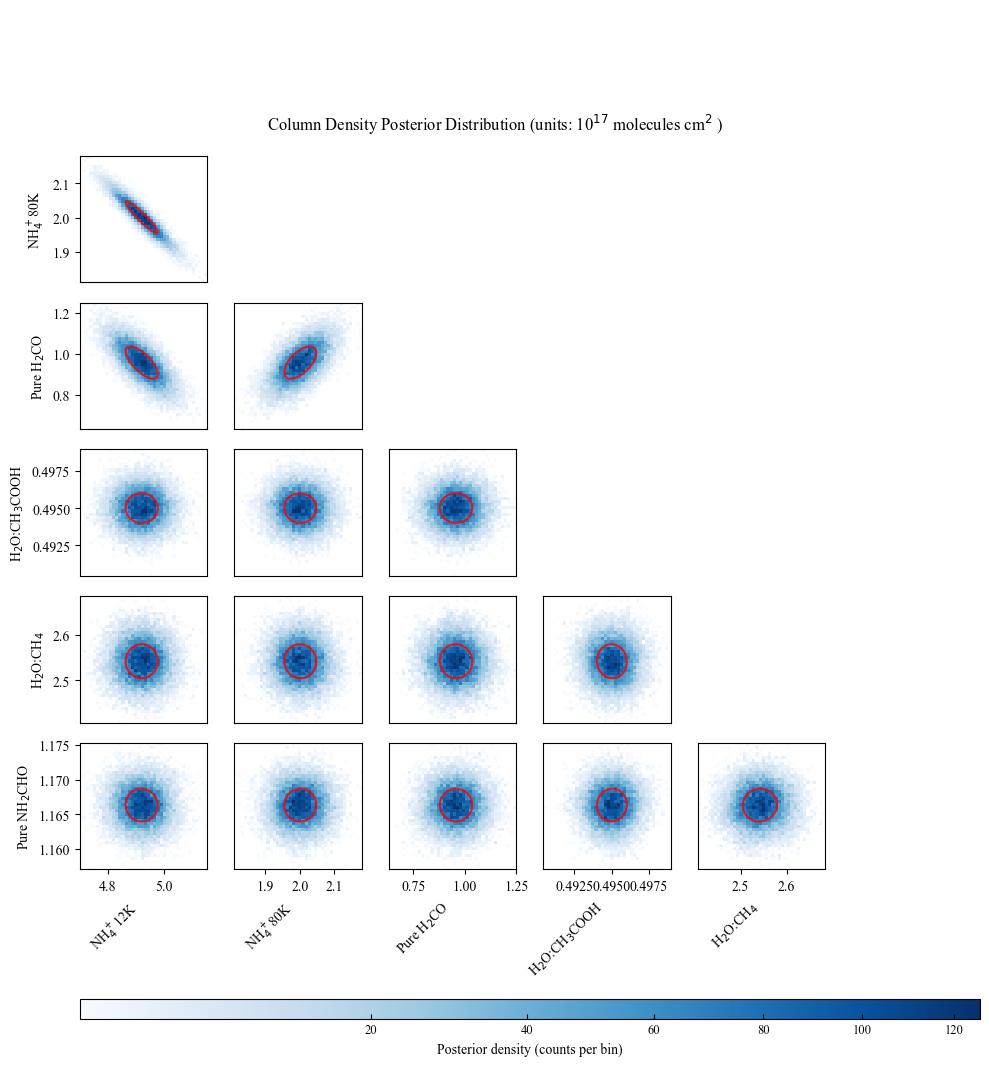}\
    \caption{Corner plot illustrating posterior distributions and confidence intervals of column densities for the various ice components fitted in the 6-\microns{8} spectral region (Step 6). The red circles indicate the 1$\sigma$ (68\%) confidence intervals, clearly displaying statistical uncertainties and correlations among the derived column densities of NH$_4^+$, H$_2$CO, CH$_4$, CH$_3$COOH, and the tentative NH$_2$CHO  ice components.}
    \label{fig_appx:Ice_fit_step6_colden}
\end{figure}

Figure~\ref{fig_appx:Ice_fit_step6} demonstrates that the absorption structure around the \microns{6.8} band is effectively reproduced by the combination of NH$_4^+$ ice components. Moreover, the global fit incorporating the remaining ice components shows excellent consistency across the entire 6-\microns{8} spectral region. The posterior distributions and confidence intervals of the derived column densities for each ice component are summarized in Figure~\ref{fig_appx:Ice_fit_step6_colden}.

However, the presence of excess absorption in the \microns{3} band indicates that a further adjustment to the initially fitted intensity of the H$_2$O ice components may be necessary, suggesting that their column densities might require refinement in a subsequent adjustment step to fully account for the interdependencies among the ice components.

\subsection{Step 7. Refinement of H$_2$O Ice Composition and Final Global Ice Analysis \label{subsec:appendix_B7}}
In this final step, we refined the previously fitted H$_2$O ice composition, with particular focus on the absorption feature around the prominent \microns{3} band. To achieve this, we first subtracted the combined absorption contributions of all ice components analyzed in earlier steps, excluding H$_2$O, thereby isolating the spectral region dominated by water-ice absorption. Then we performed an additional MCMC fitting specifically targeting this residual absorption region to precisely recalibrate the intensities of the pure amorphous (10~K) and crystalline (160~K) H$_2$O ice profiles.

\begin{figure}[hp!]
    \centering
    \includegraphics[scale=0.23]{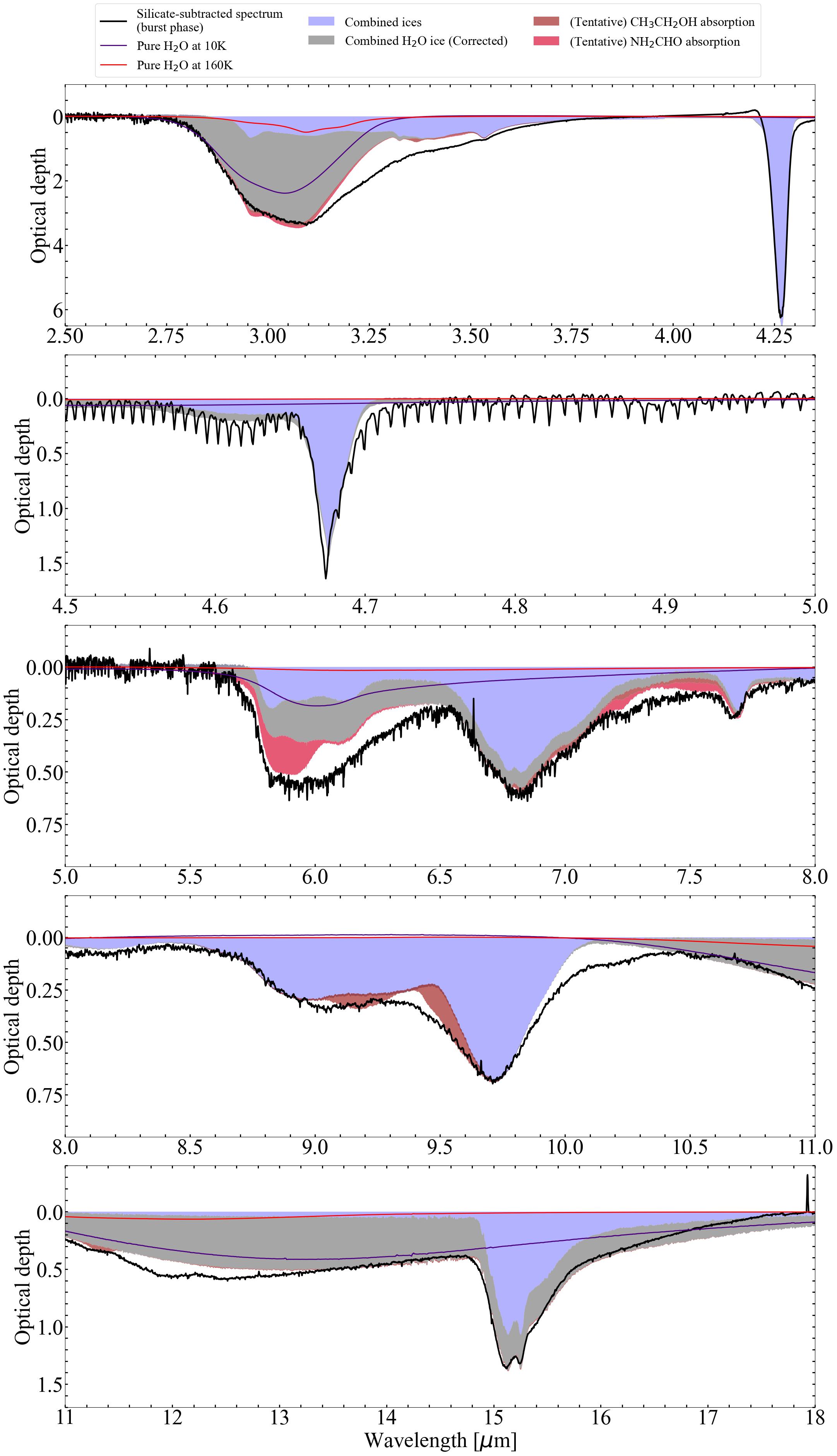}\
    \caption{Seventh and final step of the global ice-fitting procedure for EC~53, aimed at refining the absorption intensities of the previously fitted H$_2$O-based ice components, particularly around the prominent \microns{3} band. The combined ice absorption profile (blue-shaded region), derived from Step 6, incorporates all previously analyzed ice components (H$_2$O, NH$_3$, CO$_2$, CO, OCN$^-$, and CH$_3$OH-based ices). The refined fit features adjusted contributions from pure amorphous (10~K, purple line) and crystalline (160~K, red line) H$_2$O ice profiles, significantly reducing residual absorption. This step effectively resolves the previously identified discrepancies in the \microns{3} band due to overlapping ice absorptions from various minor species, yielding a final, optimized global ice composition.}
    \label{fig_appx:Ice_fit_step7}
\end{figure}

The refined MCMC fitting reduced the initially determined absorption intensities, resulting in a 5.3\% decrease for the amorphous 10~K component and a 34.4\% decrease for the crystalline 160~K component. In particular, as shown in Figure~\ref{fig_appx:Ice_fit_step7}, the crystalline H$_2$O ice absorption at 160~K was substantially adjusted downward due to the significant overlap with the absorption features of other minor ice species in the red-side region of the \microns{3} band.

\begin{figure}[hp!]
    \centering
    \includegraphics[scale=0.5]{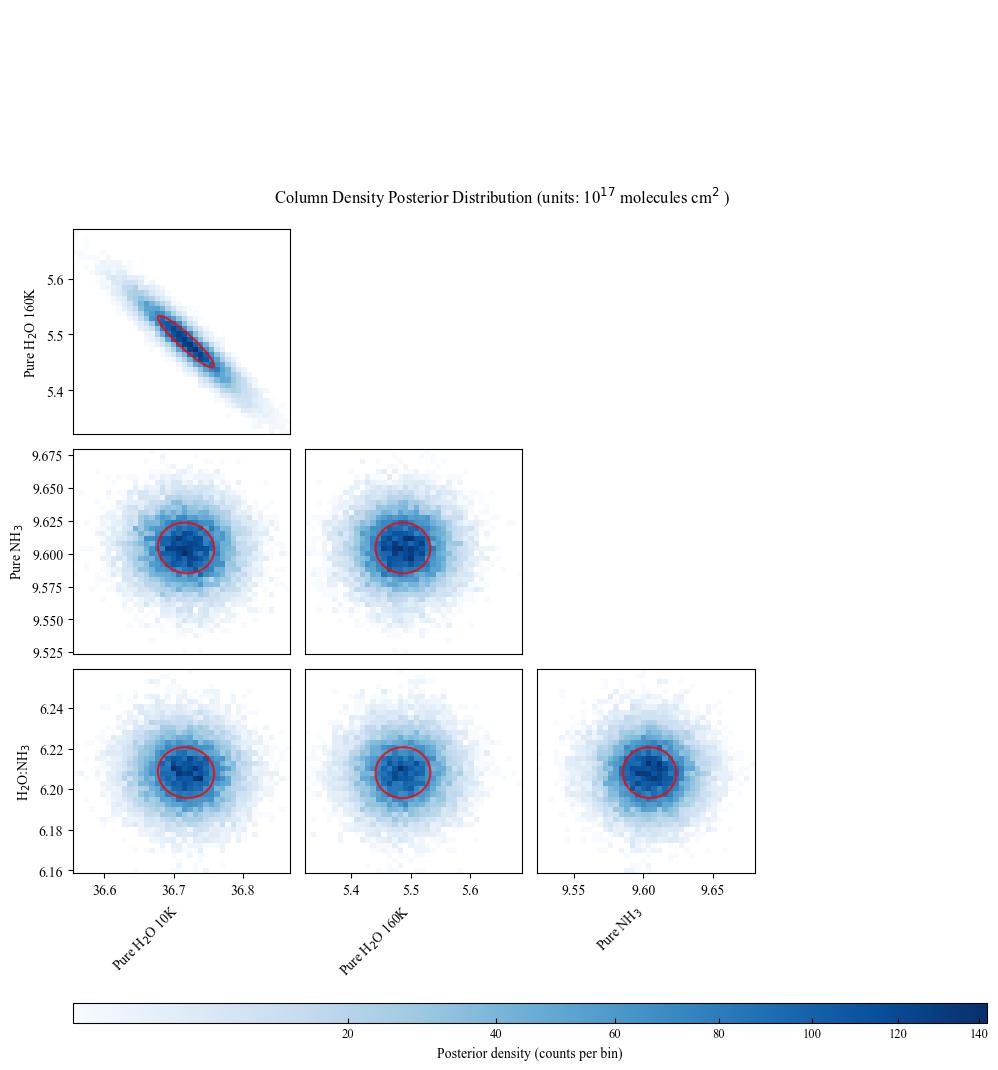}\
    \caption{Corner plot illustrating posterior distributions and refined column density confidence intervals for the primary ice components (H$_2$O- and NH$_3$-based) around the corrected absorption at the \microns{3} band. Red circles represent the 1$\sigma$ (68\%) confidence intervals, highlighting correlations and improved statistical constraints following the final adjustment performed in Step 7.}
    \label{fig_appx:Ice_fit_step7_colden}
\end{figure}

Despite this notable intensity adjustment, the resulting posterior distributions and confidence intervals of column densities (Figure~\ref{fig_appx:Ice_fit_step7_colden}) confirm that the refined combination of H$_2$O ice components accurately reproduces the observed absorption feature near \microns{3}. Figure~\ref{fig:Ice_composition_all} provides a comprehensive overview of the final global ice composition resulting from this refinement, clearly delineating the contributions of each fitted ice species across the full spectral range.

\clearpage
\setcounter{figure}{0}
\section{Possibility of the crystalline silicate absorption in the 9 to \microns{10} region \label{sec:appendix_C}}
The global ice decomposition described in Section~\ref{sec:results_global} was initially conducted on silicate-subtracted spectra, which employed amorphous silicate absorption profiles.
However, we found systematic discrepancies in the spectral region between 9 and \microns{10} that could not be fully reproduced by ice components alone (Figure~\ref{fig:Ice_composition_all}, gray-shaded region).
Such discrepancies suggest the presence of crystalline silicate absorption features, which produce characteristic narrow bands in this wavelength range due to their distinct stretching modes \citep{Molster2005,Henning2010}.

To address these spectral mismatches, we synthesized absorption profiles for crystalline forsterite and enstatite, modeled in a manner similar to that of the amorphous silicate profiles.
We fitted these crystalline silicate profiles simultaneously with previously determined ice compositions, which were dominated by NH$_3$- and CH$_3$OH-based mixtures.

\begin{figure}[hp!]
    \centering
    \includegraphics[scale=0.5]{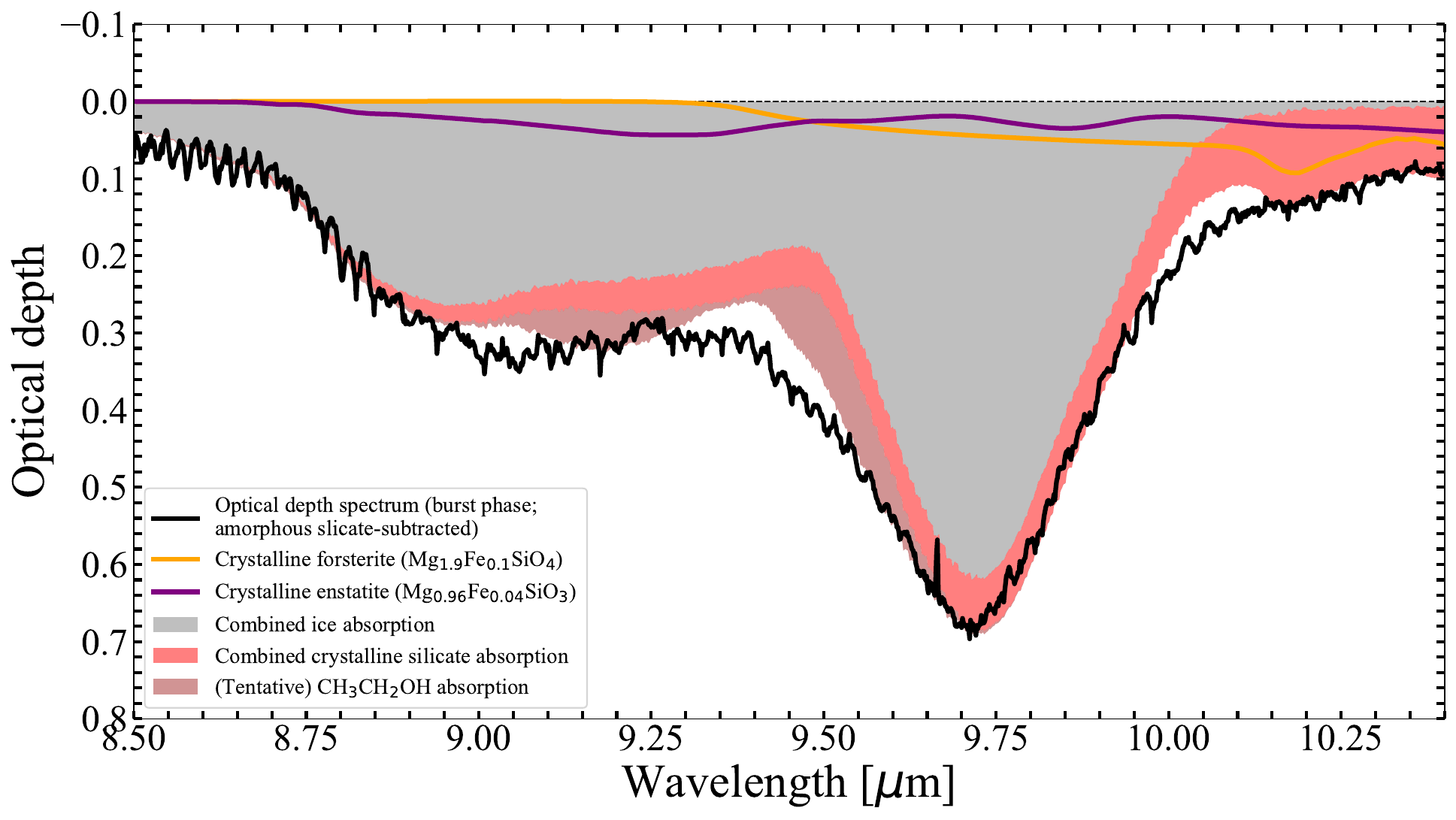}\
    \caption{Global ice decomposition applied to the optical depth spectrum of EC 53 (burst phase) in the 8.5-\microns{10.5} region, highlighting contributions from crystalline silicate absorption components. We incorporated silicate profiles for the crystalline forsterite and enstatite absorption bands, based primarily on ice composition results derived from NH$_3$- and CH$_3$OH-rich mixtures. The combined ice absorption components are shown as a gray-shaded area, while the additional absorption contributed by crystalline silicates is represented by the red-shaded area. The resulting combination improves the fit in this spectral region, especially near \microns{10.2}. Additionally, a tentative absorption contribution from ethanol (CH$_3$CH$_2$OH) at approximately \microns{9.17} is indicated to further refine the spectral fit.}
    \label{fig_appx:Cryst_Silicate}
\end{figure}

As shown in Figure~\ref{fig_appx:Cryst_Silicate}, incorporating crystalline silicates (red-shaded area) improves the spectral fit between 9 and \microns{10} compared to the fit that includes only ice components (gray-shaded area in the third panel of Figure~\ref{fig:Ice_composition_all}). 
In particular, the fitted crystalline forsterite component reproduces the absorption extending beyond \microns{9.7}, which cannot be explained by CH$_3$OH-based ice absorption alone.
This result suggests that a fraction of crystalline silicates may reside in the cold, dense disk midplane, likely transported downward after crystallization at the hot inner disk surface during accretion bursts (Paper I).

\end{appendix}

\bibliography{ref}{}
\bibliographystyle{aasjournal}



\end{document}